%% file: main.tex
\pdfoutput=1
\documentclass[a4paper,11pt]{article}

\usepackage{graphicx} 
\usepackage[margin=1in]{geometry}
\usepackage{algorithm}
\usepackage[noend]{algpseudocode}
\usepackage{amsthm}
\usepackage{amssymb,amsmath}
\usepackage{thmtools}
\usepackage{mathtools}
\usepackage[colorlinks=true, allcolors=blue]{hyperref}
\usepackage[capitalize]{cleveref}
\usepackage{placeins}
\usepackage{subcaption}
\usepackage{booktabs}

\usepackage{newtxtext}
\usepackage{newtxmath}
\usepackage{microtype}

\usepackage{enumitem}
\setlist[1]{labelindent=\parindent}
\setlist[enumerate]{label=(\arabic*)}
\setlist[itemize]{noitemsep}
\usepackage{orcidlink}
\usepackage{authblk}

\author[1]{Lorraine A. K. Ayad \orcidlink{0000-0003-0846-2616}}
\author[2]{Grigorios Loukides \orcidlink{0000-0003-0888-5061}}
\author[3]{Solon P. Pissis \orcidlink{0000-0002-1445-1932}}
\affil[1]{Brunel University of London, United Kingdom}
\affil[2]{King's College London, United Kingdom}
\affil[3]{The Cyprus Institute, Nicosia, Cyprus}

\declaretheorem[numberwithin=section]{theorem}

\declaretheorem[sibling=theorem]{lemma}

\declaretheorem[sibling=theorem]{fact}

\declaretheorem[sibling=theorem]{example}

\declaretheorem[sibling=theorem]{proposition}
\newtheorem{problem}{Problem}



\usepackage{xspace}
\usepackage{tikz}
\usetikzlibrary{arrows.meta}
\usetikzlibrary{decorations.pathreplacing}

\def\dd{\mathinner{.\,.}}

\newcommand{\cO}{\mathcal{O}}
\newcommand{\LCP}{\operatorname{LCP}}
\newcommand{\MS}{\textsf{MS}\xspace}

\title{Location-Aware Matching Statistics}

\date{\vspace{-.5cm}}

\begin{document}

\maketitle

\begin{abstract}
Edit distance is the standard measure of string similarity, but its
quadratic-time complexity is prohibitive for long strings.
Popular linear-time alternatives, such as the $q$-gram distance, or the average common substring measure (ACS) that is based on 
matching statistics (MS), consider the substrings shared by the two input strings but
not \emph{where} they occur in each string. Yet in many real-world applications, e.g., when comparing document versions or co-recorded time series, a shared substring is really relevant \emph{only} if its occurrences are close in position.

We introduce Location-Aware Matching Statistics (LAMAS). Given strings
$X$ and $Y$ and a displacement tolerance $d$, LAMAS reports, for every
position $i$ of $Y$, the longest substring starting at $i$ that occurs
in $X$ at a position $j$ with $|j-i|\le d$, together with the
displacement $(j-i)$. MS is the special case of LAMAS with unbounded $d$.
Our main result is an algorithm that solves LAMAS in $\cO(n)$ time and
space for strings of length at most $n$ and any $d$; this is \emph{optimal} 
and matches the complexity of MS. The algorithm reduces LAMAS to
predecessor and successor queries over a sliding window of suffix
ranks and answers them in linear total time via a window decomposition
and interval union-find. 

Experiments on $20$ real datasets and synthetic data show that LAMAS  
substantially improves over MS and $q$-gram distance in similarity-based classification, top-$K$ retrieval, and clustering. It is competitive with edit distance and a constrained dynamic time warping variant in classification and generally  outperforms a state-of-the-art method for top-$K$ retrieval under edit distance. Finally, our implementation of the main algorithm for LAMAS scales linearly in $n$ and does not slow down as $d$ grows.   
\end{abstract}

\thispagestyle{empty}
\clearpage
\pagenumbering{arabic}
\setcounter{page}{1}

\section{Introduction}\label{sec:intro}

Strings (i.e., sequences of elements over an alphabet) are a fundamental data type in domains ranging from
databases~\cite{DBLP:books/mg/SKS20} and information
retrieval~\cite{DBLP:books/aw/Baeza-YatesR2011} to
bioinformatics~\cite{DBLP:books/cu/Gusfield1997}.  
For example, time 
series are often discretized into
strings over a small alphabet~\cite{SAXSurvey,DBLP:journals/datamine/LinKWL07}, 
and DNA sequences are naturally represented as  strings over the four-letter alphabet of
nucleotides. Measuring the similarity of two strings is central to many 
real-world applications, such as top-$K$ similarity
search~\cite{DBLP:conf/kdd/Zhang020,DBLP:journals/tkde/LoukidesPS23},
record linkage~\cite{DBLP:books/sp/ChristenRS20},
clustering~\cite{yang2014molecular}, and
classification~\cite{DBLP:books/daglib/0021593}.  

Perhaps the most common way to compare two strings is to compute their \emph{edit distance} (ED), also known as the \emph{Levenshtein distance}~\cite{Levenshtein1966}: the minimum number of character
insertions, deletions, and substitutions required to transform
one string into another.
For two strings, each of length at most $n$, the classical dynamic-programming
algorithm computes their ED in $\cO(n^2)$ time~\cite{DBLP:journals/jacm/WagnerF74}. 
Unfortunately, this upper bound has been improved only by polylogarithmic factors~\cite{DBLP:journals/siamcomp/CrochemoreLZ03,DBLP:journals/tcs/BilleF08,DBLP:journals/dam/Grabowski16}, while conditional lower bounds suggest that no truly subquadratic algorithm exists~\cite{DBLP:journals/siamcomp/BackursI18,DBLP:conf/focs/AbboudBW15,DBLP:conf/focs/BringmannK15,DBLP:conf/stoc/AbboudHWW16}.

To bypass this quadratic barrier, many efficiently computable measures have been developed that compare the constituent patterns of the two strings, which we denote by $X$ and $Y$. The widely used \emph{$q$-gram distance}~\cite{DBLP:journals/tcs/Ukkonen92} compares the frequencies of length-$q$ substrings of $X$ and $Y$, while the \emph{average common substring} (ACS) measure averages, over all positions of each string, the length of the longest substring starting at that position 
that also occurs in the other string~\cite{DBLP:journals/jcb/UlitskyBTC06}. 
Unlike ED, both measures
can be computed in $\cO(n)$
time (cf.~\cite{DBLP:journals/tcs/Ukkonen92,DBLP:conf/focs/ChangL90}).
However, they capture only \emph{which} patterns the two strings share,
not \emph{where} these patterns occur: an occurrence in $X$ contributes
equally to ACS whether it lies close to the corresponding
position of $Y$ or far away from it. 

\subsection{Motivation}

In many applications, positions in strings $X$ and $Y$ are comparable,
and the relevance of a shared pattern depends on \emph{where} its occurrences are located. 
For example, when comparing two versions of a document to identify unchanged content~\cite{DBLP:journals/spe/MillerM85}, a shared phrase occurring near the same position is more informative than one appearing only in an unrelated section. Yet exact positional agreement is too restrictive, since insertions and deletions shift subsequent text.

A similar issue arises for time series discretized into 
strings. Consider two 
stock-price series recorded over the same period
and encoded as strings $X$ and $Y$. A shared rise-and-fall pattern in two stock-price series indicates similar behavior more strongly when it occurs at approximately the same time. 
In fact, allowing limited temporal shifts is well established in
time-series matching. For example, the well-known \emph{dynamic time warping}
(DTW)~\cite{rw:Keogh02} measure aligns observations at different positions, while banded DTW restricts the alignment path 
to a band of radius $d$ around the diagonal~\cite{SakoeChiba78}, like the diagonal-band version of ED~\cite{DBLP:journals/iandc/Ukkonen85}. As for ED, the textbook $\cO(n^2)$-time algorithm for DTW~\cite{SakoeChiba78} has been improved only by a sub-logarithmic factor~\cite{DBLP:journals/talg/GoldS18} and is unlikely to admit significant improvements~\cite{DBLP:conf/focs/AbboudBW15,DBLP:conf/focs/BringmannK15}, whereas banded DTW takes $\cO(nd)$ time. 
Positional constraints are also used in time-series database querying~\cite{FuKeoghLauRatanamahatana2005} and in string indexing, where pattern searches are confined to a specified text interval~\cite{DBLP:conf/latin/MakinenN06}.

The following folklore observation
underlies many algorithms for computing ED and justifies the above intuition; see \Cref{sec:prelim} for the definition of alignments.

\begin{lemma}[\cite{DBLP:journals/iandc/Ukkonen85}]
\label{lem:mot}
Fix any alignment of $X$ and $Y$ of cost at most $k$. If $X[j]$ is
aligned with $Y[i]$ (as a match or a substitution), then $|j-i|\le k$.
\end{lemma}

This leads to the \emph{central question} of this work:

\begin{center}
    \fcolorbox{gray!50}{gray!15}{%
        \begin{minipage}{0.9\linewidth}
            \centering
            \emph{Can we design a string-similarity measure based on shared patterns that
            takes the positions of these patterns into consideration and is still computable
            in linear time?}
        \end{minipage}%
    }
\end{center}

\subsection{Our approach}

To answer this question, we build on the \emph{matching statistics} (MS) primitive~\cite{DBLP:conf/focs/ChangL90,DBLP:books/cu/Gusfield1997,DBLP:journals/tcs/LiptakMP26}, which underlies ACS, rather than on $q$-grams, because MS \emph{does not fix the length} of the patterns: it maximizes the match length at every position of $Y$, capturing arbitrarily long matches. 
Formally, given two strings $X=X[0\dd |X|)$ and $Y=Y[0\dd |Y|)$, the  matching statistics of $Y$ with respect to $X$ is the array $\MS_X=\MS_X[0\dd |Y|)$ defined, for all $i\in[0\dd |Y|)$, by
\begin{equation}\label{eq:ms}
\MS_X[i]=\max_{0\le j<|X|}\LCP\bigl(X[j\dd |X|),Y[i\dd |Y|)\bigr),
\end{equation}
where $\LCP(U,V)$ denotes the length of the longest common prefix (LCP) of two
strings $U$ and $V$.  
Intuitively, $\MS_X[i]$ is the length of the longest substring of $Y$ that
starts at position $i$ and occurs \emph{anywhere} in $X$.
The definition of the array $\MS_Y$ is analogous. Both arrays can be
computed in $\cO(|X|+|Y|)$ time~\cite{DBLP:conf/focs/ChangL90}. The ACS measure for $X$ and $Y$ is defined as:\footnote{We omit logarithmic correction terms from this formula for simplicity~\cite{DBLP:journals/jcb/UlitskyBTC06}.}
\[
\operatorname{ACS}(X,Y)=\frac{1}{2}\Bigl(\frac{1}{|Y|}\sum_{i=0}^{|Y|-1}\MS_X[i]+\frac{1}{|X|}\sum_{i=0}^{|X|-1}\MS_Y[i]\Bigr).
\]
The unrestricted choice of $j$ in \Cref{eq:ms} allows a longer
distant match to be selected instead of a shorter nearby
one,  which is
undesirable when positions in the two strings are comparable.
Moreover, simply discarding distant
matches from the output of MS does not recover the best nearby match,
since shorter alternatives are not reported. 

We therefore introduce \emph{Location-Aware Matching Statistics}
(LAMAS), which restricts the starting position of each match to lie
within a \emph{displacement tolerance} $d$ of the query position: for each position $i$ of $Y$, LAMAS finds the longest substring of $Y$ starting at position $i$ that
occurs in $X$ starting at a position $j$ with $|j-i|\le d$, and returns its
length $\ell_i$ and displacement $\delta_i=j-i$.

\begin{problem}[LAMAS]\label{problem:LAMAS}
Given two strings $X$ and $Y$, each of length at most $n$, and an
integer $d\ge 0$, output, for every $i\in[0\dd |Y|)$, a pair
$(\ell_i,\delta_i)$, where
\[
\ell_i=\max_{\substack{0\le j<|X|\\ |j-i|\le d}}\LCP\bigl(X[j\dd|X|),Y[i\dd|Y|)\bigr)
\]
and $\delta_i=j-i$ for a position $j$ attaining this maximum (ties are
broken arbitrarily). If no position $j\in[0\dd |X|)$ satisfies
$|j-i|\le d$, then $\ell_i=0$ and $\delta_i$ is undefined.
\end{problem}

As for MS, we denote the array of lengths by $\textsf{LMS}_X$,
i.e., $\textsf{LMS}_X[i]=\ell_i$, and define $\textsf{LMS}_Y$
analogously.

\begin{example}
\label{ex:ms-acs-lamas}
Let $X=\mathtt{abcdef}$ and $Y=\mathtt{defabc}$.
The longest prefix of $Y[0\dd 6)$ occurring in $X$ is $\mathtt{def}$, at $j=3$, so $\MS_X[0]=3$; similarly, $\MS_X=\MS_Y=(3,2,1,3,2,1)$. For LAMAS with $d=2$, the admissible positions for $i=0$ are $j\in\{0,1,2\}$, none of which starts with $Y[0]=\mathtt{d}$, so $\textsf{LMS}_X[0]=0$. Likewise, every letter of $Y$ occurs in $X$ only at displacement $\pm3$, so $\textsf{LMS}_X=(0,0,0,0,0,0)$, and $(\ell_i,\delta_i)=(0,0)$ for every $i$ is a valid output.
\end{example}

The tolerance $d$ can be set from domain knowledge. In time series, it
is the largest acceptable delay between corresponding events, as for the
warping window of banded DTW~\cite{SakoeChiba78}. When the strings are expected to
be within ED $k$, as is the case for closely related DNA sequences,
setting $d=k$ ensures that every exact match of an alignment of cost at
most $k$ starts at an admissible position (\Cref{lem:mot}). When no such
knowledge is available, $d$ can be estimated from a sample of the data 
(e.g., from the EDs of a few sampled pairs), or tuned on labeled
validation data, as in our experiments (see \Cref{sec:pair-classification}). 

\Cref{ex:prices} shows, on real stock prices, where the timing of
price movements matters~\cite{MacKinlay1997Stock}, that ignoring
positions can invert a similarity ranking, and that LAMAS corrects it.

\begin{figure}[t]
    \centering
    \includegraphics[width=.55\linewidth]{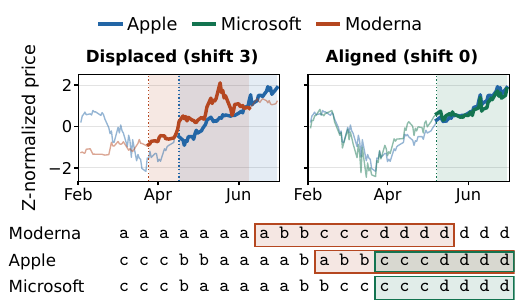}
    \caption{Shared patterns in stock-price time series.}
    \label{fig:stock-example}
\end{figure}

\begin{example} \label{ex:prices}
We discretize the daily adjusted closing prices of Apple, Microsoft, and
Moderna stocks from 3 February to 30 June 2020~\cite{YahooFinanceStock2020}
using SAX~\cite{Lin2003SAXStock} into strings of length $n=20$ over
$\sigma=4$ letters, so that each letter spans about one trading week
(\Cref{fig:stock-example}). Apple and Moderna share the long pattern
\texttt{abbcccdddd}, but its occurrences are three positions (about three weeks) apart, 
whereas Apple and Microsoft share the shorter pattern \texttt{cccdddd}
at the same position. Hence, ACS ranks Moderna more similar to Apple 
than Microsoft ($4.35$ vs.\ $3.85$). Replacing MS in ACS  with LAMAS at $d=1$ excludes the displaced match and reverses the ranking
($1.425$ vs.\ $3.1$), as does using LAMAS at $d=0$ or $d=2$. This ranking agrees with the Pearson correlation of daily log returns, an established measure of stock co-movement~\cite{Plerou2002}: $0.9374$ for Apple--Microsoft versus only $0.0027$ for Apple--Moderna.\end{example}

\Cref{ex:acs-ed} shows that ACS based on MS can rate two strings as highly
similar when their ED is maximal, whereas LAMAS avoids this when $d$ excludes the distant matches. 

\begin{example}\label{ex:acs-ed}
Let $n$ be divisible by $3$, and consider
\[X=\mathtt{a}^{n/3}\mathtt{c}^{n/3}\mathtt{g}^{n/3}
\quad\text{and}\quad
Y=\mathtt{g}^{n/3}\mathtt{t}^{n/3}\mathtt{a}^{n/3},
\]
where $U^k$ denotes the concatenation of $k$ copies of string $U$.
The ED of $X$ and $Y$ is $n$, the maximum possible. 
Nevertheless, MS matches the shared blocks
$\mathtt{a}^{n/3}$ and $\mathtt{g}^{n/3}$, so
ACS rates $X$ and $Y$ as highly
similar. In contrast, all shared letters lie more than $n/3$ positions
apart, so for every $d\le n/3$, we have
$\textsf{LMS}_X[i]=\textsf{LMS}_Y[i]=0$ for all $i$, and so LAMAS reports no
similarity at all, agreeing with the ED.
\end{example}

\subsection{Contributions}

In addition to introducing the LAMAS problem, our work makes the following contributions. 

\paragraph{A time-optimal algorithm for LAMAS.}
Standard string indexes support
longest common extension (LCE) queries in $\cO(1)$ time after
$\cO(n)$-time preprocessing~\cite{DBLP:journals/jacm/KarkkainenSB06,DBLP:conf/cpm/KasaiLAAP01,DBLP:conf/latin/BenderF00}.
Thus, computing LAMAS by checking all $\cO(d)$ admissible positions in one string for each starting position   
in the other takes $\cO(nd)$ total time. Our central contribution is
an $\cO(n)$-time and $\cO(n)$-space
algorithm for LAMAS. 

We observe that, for each starting position in one string, at most  two admissible suffixes of the other need to be checked: the lexicographically nearest admissible suffix on either side of the query suffix (formally, the predecessor and successor of $Y[i\dd|Y|)$ among suffixes of $X$ starting at positions $j$ with $|j-i|\le d$). 
Maintaining their ranks in a balanced search tree as the admissible window slides yields a simple $\cO(n\log d)$-time  
algorithm. We improve this algorithm by: 

\begin{itemize}
    \item \emph{Reducing the search space.} Each group of nearby queries involves only a small subset of the globally ordered suffixes. We exploit this locality by partitioning the queries into blocks and replacing global suffix ranks with local ranks, which preserves the answers while shrinking the search universe. This yields $\cO(n\log\log d)$ time, and $\cO(n)$ time when $d=\log^{\cO(1)}n$.

    \item \emph{Exploiting the window structure.} Consecutive query windows overlap substantially and change predictably. Decomposing each window into a fixed part and two parts that shrink under suitable query orders lets us replace general-purpose dynamic search by sorted scans and \textit{interval union-find}~\cite{DBLP:journals/jcss/GabowT85}, which process all queries in $\cO(n)$ total time, independently of $d$.
\end{itemize}

Thus, we achieve optimal $\cO(n)$ time and
$\cO(n)$ space, matching the asymptotic
complexity of classical MS \emph{despite the
additional displacement constraint} and answering  our central question in the affirmative.

\paragraph{Experimental evaluation.}
We evaluate LAMAS on $20$ real datasets
and synthetic data, comparing it with MS, $q$-gram
distance, ED, banded DTW~\cite{SakoeChiba78}, and 
MinSearch~\cite{DBLP:conf/kdd/Zhang020}, a state-of-the-art
method for top-$K$ retrieval under ED.   
We consider three common similarity-based tasks:

\begin{enumerate}
    \item \emph{Pair classification.}
    LAMAS outperforms MS on all $15$ tested datasets,
    achieving a mean $F_1$ of $0.7644$, which is 
    $9.2\%$ and $8.7\%$ higher than MS and
    $q$-gram distance, respectively. 
    It also achieves effectiveness comparable to ED and banded DTW~\cite{SakoeChiba78}.  

    \item \emph{Top-$K$ retrieval.}
    LAMAS outperforms MS and $q$-gram distance
    on all $15$ tested datasets; its mean
$F_1$@10~\cite{DBLP:books/daglib/0021593} is $0.7516$; $74.9\%$, 
$57\%$, and $15\%$ higher than MS,
$q$-gram distance, and MinSearch, respectively.
It also achieves the highest mean retrieval
effectiveness for $K\geq5$.

    \item \emph{Clustering.} LAMAS outperforms MS and $q$-gram distance
    on four quality measures across two
    real datasets, improving ARI~\cite{HubertArabie1985}
     by up to $146\%$ and $154\%$, respectively.
     It also outperforms ED by up to $96\%$ in ARI.
\end{enumerate}

\paragraph{Efficiency.}
Our LAMAS algorithm takes a similar time with MS and less memory on DNA sequencing datasets. It also scales linearly in $n$ and does not slow down as $d$ grows. 

\subsection{Paper organization}
\Cref{sec:related} discusses related work and  \Cref{sec:prelim} the preliminaries. \Cref{sec:algo} presents our LAMAS algorithm and   \Cref{sec:experiments} our experiments.  \Cref{sec:conclusion} concludes the paper. 

\section{Related Work}\label{sec:related}

\paragraph{Classical string-similarity measures.}
Classical string-similarity measures include Hamming distance,
ED, longest common subsequence (LCS), and
their variants; see~\cite{rw:Navarro01} for a survey.
Hamming distance counts positional mismatches, while ED allows insertions, deletions, and substitutions.
Damerau--Levenshtein distance additionally allows adjacent
transpositions~\cite{rw:Damerau64}.
LCS permits only insertions and deletions (gaps)~\cite{rw:Hirschberg75},
whereas longest common substring~\cite{DBLP:journals/talg/CharalampopoulosKRP26} requires contiguity. Unlike these, LAMAS records, at each position,
the longest exact match within a bounded displacement,
providing an efficient, location-aware, proxy for ED-based
similarity.

\paragraph{Edit-distance embeddings and learned similarities.}
Metric embeddings map strings into Hamming space
while approximately preserving ED~\cite{ChakrabortyGK16,DBLP:conf/icalp/CharikarGKK18}. EmbedJoin uses such embeddings for edit-similarity joins
\cite{DBLP:conf/kdd/Zhang017}.
Learning-based methods instead learn edit costs
\cite{DBLP:journals/pami/RistadY98} or similarity representations
\cite{DBLP:conf/sigir/DaiYZW0C20,
DBLP:conf/acl-spnlp/LibovickyF22,
DBLP:journals/kbs/LiuWYH25}.
Unlike these methods, LAMAS requires no training, and,
unlike the (randomized) embeddings, it is deterministic; however,
it offers no formal ED approximation guarantees.

\paragraph{Alignment-free string measures.}
Alignment-free measures compare strings through shared or absent
patterns. The $q$-gram distance compares substring frequencies
\cite{DBLP:journals/tcs/Ukkonen92}, while its variant, threshold $q$-gram distance, caps
multiplicities to reduce space \cite{rw:CenzatoFLM25}. $q$-gram-based distances are used in several bioinformatics tools, e.g., in
\cite{rw:Grossi16,rw:AyadPissis17}.
Minimal-absent-word (MAW) measures compare absent patterns
\cite{rw:ChairungseeCrochemore12,rw:Charalampopoulos18}.
Matching statistics (MS)~\cite{rw:ChangLawler94} is a widely used primitive; see \cite{DBLP:journals/tcs/LiptakMP26} for a recent survey. It also underpins software tools such as those in \cite{rw:AhmedRKSG20,rw:RossiOLGB22}. LAMAS generalizes MS by bounding the displacement between matching starting positions, which benefits multiple applications as we show experimentally, and it can be computed within the same (optimal) time complexity.

\paragraph{Distances for time series.}
Time series can be compared using elastic measures, e.g., DTW~\cite{rw:Keogh02}, and LCSS~\cite{rw:VlachosGK02}; 
see~\cite{rw:HolderMB24} for a survey.
Order-preserving matching statistics (OPMS) instead captures
similarity in relative value order, enabling trend-based
comparison~\cite{rw:LiZLPM26}; this is orthogonal to LAMAS.
Dynamic sub-sequence warping (DSSW) compares extracted
numerical features~\cite{rw:ZhouHD26}. Another popular way to compare time series is to first discretize them into strings, e.g., using
SAX (symbolic aggregate approximation )~\cite{DBLP:journals/datamine/LinKWL07,Lin2003SAXStock}. This reduces dimensionality
and storage while enabling string-based processing.
LAMAS operates on symbolic sequences, and,
unlike these elastic measures, which require quadratic time to be computed in general, it can be computed in the optimal $\cO(n)$ time.

\section{Preliminaries}\label{sec:prelim}

\paragraph{Model of computation.}
Throughout this paper, we assume the standard word RAM model with
machine words of $w=\Omega(\log n)$ bits, where $n$ is the total size of the input.

\paragraph{Strings.}
For integers $i$ and $j$, we write $[i\dd j]$ for
$\{i,\ldots,j\}$ and $[i\dd j)$ for $\{i,\ldots,j-1\}$; both are empty if $i>j$, and $[i\dd j)=\emptyset$ also if $i=j$.
An \emph{alphabet} $\Sigma$ is a finite totally ordered set, whose
elements are called \emph{letters}. We assume an integer alphabet
$\Sigma=[0\dd \sigma)$ with $\sigma=n^{\cO(1)}$, so that letters fit in
machine words. A \emph{string} $X=X[0]X[1]\cdots X[|X|-1]$
over $\Sigma$ is a sequence of $|X|$ letters, where $|X|$ is its
\emph{length}; the \emph{empty string} $\varepsilon$ has length $0$.
For $0\le i\le j\le |X|$, the \emph{fragment} $X[i\dd j)$ is the string
$X[i]X[i+1]\cdots X[j-1]$ (it is empty if $i=j$), and we write
$X[i\dd j]$ for $X[i\dd j+1)$. A fragment $X[0\dd j)$ is a
\emph{prefix} of $X$ and a fragment $X[i\dd |X|)$ is a \emph{suffix}
of $X$; a prefix or suffix is \emph{proper} if it differs from $X$.
The concatenation of two strings $U$ and $V$ is denoted by $UV$ 
A string
$W$ \emph{occurs} in $X$ at position $i$ if $W=X[i\dd i+|W|)$.

\paragraph{Lexicographic order and longest common prefixes.}
The \emph{longest common prefix} (LCP) of two strings $U$ and $V$ is the
longest string that is a prefix of both; we denote its length by $\LCP(U,V)$. 
The
\emph{lexicographic order} $\prec$ on strings is defined by $U\prec V$
if and only if either $U$ is a proper prefix of $V$, or
$\ell=\LCP(U,V)<\min(|U|,|V|)$ and $U[\ell]<V[\ell]$. We write
$U\preceq V$ if $U\prec V$ or $U=V$.
We use \Cref{fact:lcp-min} extensively.

\begin{fact}[\cite{DBLP:journals/siamcomp/ManberM93}]
Let $U\preceq V\preceq W$ be strings. Then
$\LCP(U,W)=\min\{\LCP(U,V),\LCP(V,W)\}$.\label{fact:lcp-min}
\end{fact}

\paragraph{Edit distance and alignments.}
An \emph{alignment} of two strings $X$ and $Y$ is a sequence
$\bigl((u_1,v_1),\allowbreak\ldots,(u_z,v_z)\bigr)$ of \emph{columns}
$(u_t,v_t)\in(\Sigma\cup\{\varepsilon\})^2\setminus\{(\varepsilon,\varepsilon)\}$
such that $u_1u_2\cdots u_z=X$ and $v_1v_2\cdots v_z=Y$. A column
$(u_t,v_t)$ is a \emph{match} if $u_t=v_t$, a \emph{substitution} if
$u_t\ne v_t$ and $u_t,v_t\in\Sigma$, a \emph{deletion} if
$v_t=\varepsilon$, and an \emph{insertion} if $u_t=\varepsilon$. If
$(u_t,v_t)$ is a match or a substitution, with $u_t=X[j]$ and
$v_t=Y[i]$, we say that $X[j]$ is \emph{aligned with} $Y[i]$, and
we call $j-i$ the \emph{displacement} of the column. Under
\emph{unit costs}, the cost of an alignment is its number of
non-match columns, and the \emph{edit distance} (or Levenshtein
distance~\cite{Levenshtein1966}) $\operatorname{ED}(X,Y)$ is the
minimum cost of an alignment of $X$ and $Y$.

\paragraph{$q$-gram distance.}
The $q$-gram distance
~\cite{DBLP:journals/tcs/Ukkonen92}
is the $\ell_1$ distance between the occurrence-count
vectors of all substrings of length $q$ in $X$ and $Y$.
It can be computed in linear time~\cite{DBLP:journals/tcs/Ukkonen92}. 

\paragraph{Suffix array, inverse suffix array, and LCP array.}
Let $S$ be a string of length $m$ over an integer alphabet. All
suffixes of $S$ are distinct, so the \emph{suffix array} of $S$ is
the unique permutation $\textsf{SA}_S[0\dd m)$ of $[0\dd m)$ such that  $\textsf{SA}_S[r]$ is the starting position of the
lexicographically $r$-th smallest suffix of $S$. The \emph{inverse
suffix array} $\textsf{iSA}_S[0\dd m)$ is defined by
$\textsf{iSA}_S[\textsf{SA}_S[r]]=r$; thus $\textsf{iSA}_S[i]$ is the
\emph{rank} of the suffix $S[i\dd m)$ among all suffixes of $S$, and
$S[i\dd m)\prec S[j\dd m)$ if and only if $\textsf{iSA}_S[i]<\textsf{iSA}_S[j]$.

The \emph{LCP array} $\textsf{LCP}_S[0\dd m)$ stores the LCP lengths
of lexicographically consecutive suffixes: $\textsf{LCP}_S[0]=0$ and
\[
\textsf{LCP}_S[r]=\LCP\bigl(S[\textsf{SA}_S[r-1]\dd m),\,S[\textsf{SA}_S[r]\dd m)\bigr),
\]
for $r\in[1\dd m)$.
The suffix array of a string of length $m$ over an integer alphabet
of size $m^{\cO(1)}$ can be constructed in $\cO(m)$
time~\cite{DBLP:journals/jacm/KarkkainenSB06}; the inverse suffix
array then follows in $\cO(m)$ time, and the LCP array can be
computed in $\cO(m)$ time from $S$, $\textsf{SA}_S$, and
$\textsf{iSA}_S$~\cite{DBLP:conf/cpm/KasaiLAAP01}. In
\Cref{sec:algo}, these arrays are built for the string $S=X\#Y$,
where $\#\notin\Sigma$ is a \emph{sentinel} letter, which we assume
to be lexicographically larger than every letter of $\Sigma$; the alphabet of $S$ is
thus $[0\dd \sigma]$ with $\#=\sigma$, and $|S|=|X|+|Y|+1$. The suffix
of $S$ starting at position $|X|$ is the only suffix starting with
$\#$; every other suffix of $S$ starts either in $X$ (positions
$[0\dd|X|)$) or in $Y$ (positions $[|X|+1\dd|S|)$).

\paragraph{Range minimum queries.}
Let $A[0\dd m)$ be an array. A \emph{range minimum query}
$\operatorname{RMQ}_A(i,j)$, for $0\le i\le j<m$, returns an index
$t\in[i\dd j]$ such that $A[t]=\min A[i\dd j]$ (ties are broken
arbitrarily). After preprocessing $A$ in $\cO(m)$, 
any range minimum query (RMQ) can be answered in
$\cO(1)$ time~\cite{DBLP:conf/latin/BenderF00}.

\paragraph{Longest common extension queries.}
A \emph{longest common extension} (LCE) query on a string $S$ of
length $m$ asks, for two positions $i,j\in[0\dd m)$, for the value
$\operatorname{LCE}_S(i,j)=\LCP(S[i\dd m),S[j\dd m))$. By
\Cref{fact:lcp-min} applied to the sorted sequence of suffixes, LCE
queries reduce to RMQs on the LCP array: for
$i\ne j$, letting $r=\textsf{iSA}_S[i]$ and $s=\textsf{iSA}_S[j]$ with
$r<s$ without loss of generality,
\[
\operatorname{LCE}_S(i,j)=\min\textsf{LCP}_S[r+1\dd s]=\textsf{LCP}_S[t^*],
\]
where $t^*=\operatorname{RMQ}_{\textsf{LCP}_S}(r+1,s)$,
while $\operatorname{LCE}_S(i,i)=m-i$. Hence, after an $\cO(m)$-time
preprocessing of $S$ (suffix array, inverse suffix
array, LCP array, and the RMQ structure on it), every LCE query is
answered in $\cO(1)$ time; we refer to this combination as the
\emph{LCE data structure} of $S$. For $S=X\#Y$, the LCE data structure
also answers LCP queries between a suffix of $X$ and a suffix of $Y$:
since $\#$ does not occur in $Y$, a common prefix of $S[j\dd|S|)$ and
$S[|X|+1+i\dd|S|)$ cannot extend past $\#$, so for all
$j\in[0\dd|X|)$ and $i\in[0\dd|Y|)$,
\[
\LCP\bigl(X[j\dd|X|),Y[i\dd|Y|)\bigr)=\operatorname{LCE}_S(j,\,|X|+1+i).
\]
The sentinel may change the relative order of such a pair, but never its LCP value. By \Cref{fact:lcp-min} applied to the suffixes of $S$ in suffix-array
order, among a set $R$ of ranks of suffixes of $X$ in
$\textsf{SA}_S$, the LCP with $Y[i\dd|Y|)$, whose suffix of $S$ has
rank $\rho=\textsf{iSA}_S[|X|+1+i]$, is maximized by
$\max\{r\in R: r<\rho\}$ or by $\min\{r\in R: r>\rho\}$.

\paragraph{Predecessor and successor queries.}
Given a set $K$ of integers and an integer $x$, the
\emph{predecessor} (resp.~\emph{successor}) of $x$ in $K$ is defined as
$\operatorname{pred}_K(x)=\max\{y\in K\mid y\le x\}$ (resp.~$\operatorname{succ}_K(x)=\min\{y\in K\mid y\ge x\}$); these are undefined if the corresponding set is empty. A \emph{dynamic predecessor data structure} maintains a set $K$ under
insertions and deletions and answers predecessor and successor
queries. We use three classic implementations. A balanced binary
search tree (e.g., an AVL tree) supports all operations in
$\cO(\log |K|)$ time using $\cO(|K|)$ space. For a universe
$[0\dd u)$ of integers, a van Emde Boas (vEB) tree~\cite{DBLP:journals/ipl/Boas77}
supports all operations in $\cO(\log\log u)$ time, using $\cO(u)$
space and $\cO(u)$ initialization time. A dynamic fusion
tree~\cite{DBLP:conf/focs/PatrascuT14} supports all operations in
$\cO(\log_w |K|)=\cO(\log|K|/\log w)$ time using $\cO(|K|)$ space;
in particular, in $\cO(1)$ time when $|K|=\log^{\cO(1)} n$.

\paragraph{Disjoint sets.}
A \emph{union-find} data structure maintains a partition of a ground set
$[0\dd m)$, initially consisting of singleton sets, under the operations  
$\textsf{find}(k)$, which returns the representative of the set containing
$k$, and $\textsf{union}(k,k')$, which merges the sets containing $k$ and
$k'$. In the \emph{static tree set union} problem, a rooted tree $T$ on
$[0\dd m)$ is given in advance, and every union merges the set containing a
non-root node $k$ with the set containing its parent; thus, the possible
unions are fixed before any operation is performed, whereas the order of the
operations is arbitrary. Gabow and Tarjan~\cite{DBLP:journals/jcss/GabowT85}
showed that, in this case, any sequence of $q$ operations takes $\cO(m+q)$
total time. We use the special case in which $T$ is the path where $k+1$ is
the parent of $k$, for every $k\in[0\dd m-1)$: every set is then a
contiguous interval of $[0\dd m)$, and every union merges two adjacent
intervals $[l\dd k]$ and $[k+1\dd r]$. This is known as the \emph{interval
union-find} problem.

\input{algorithm}

\section{Experimental Evaluation}\label{sec:experiments}

Our experimental evaluation addresses five questions.
\begin{itemize}
    \item[\emph{Q1}] How closely does \textsc{LAMAS} correlate with ED as sequence \emph{divergence} increases? (\Cref{accuracy}).
    \item[\emph{Q2}] How effectively does \textsc{LAMAS} classify time-series pairs? (\Cref{pair_classification}).
    \item[\emph{Q3}] How accurately does \textsc{LAMAS} recover the top-$K$ results under ED? (\Cref{top-k}).
    \item[\emph{Q4}] How effectively does \textsc{LAMAS} recover reference classes through clustering? (\Cref{clustering}).
    \item[\emph{Q5}] How efficient in terms of running time and space is our $\cO(n)$-time \textsc{LAMAS} algorithm? (\Cref{efficiency}).
\end{itemize}

We use $20$ real datasets together with synthetic datasets. We compare 
our $\cO(n)$-time \textsc{LAMAS} algorithm to: (1) our own implementation of \textsc{MS}, (2) our own implementation of $q$-gram distance, (3) the ED implementation of  
Edlib~\cite{edlib}, which uses the efficient Myers' bit-vector algorithm, (4) our $\cO(n\log d)$-time algorithm of \Cref{sec:simple}, referred to as  \textsc{LAMAS} (\textsc{Simple}), and (5) our $\cO(n\log\log d)$-time algorithm of \Cref{sec:improved}, referred to as  \textsc{LAMAS} (\textsc{vEB}).   
All \textsc{LAMAS} implementations and MS use SDSL-lite~\cite{sdsl}. For the classification experiments, we also compare against banded DTW~\cite{SakoeChiba78}, which bounds the displacement between aligned positions and is thus a natural baseline.
For the top-$K$ retrieval experiments, we 
additionally compare against an adapted version of 
MinSearch~\cite{DBLP:conf/kdd/Zhang020}. We do not report results for the edit-distance embedding method of~\cite{ChakrabortyGK16}, as it underperformed LAMAS consistently in our setting. For example, in the experiment of \Cref{accuracy}, its mean Pearson correlation 
with ED was $70\%$ lower than that of the worst competitor.  

All experiments ran on an Intel(R) Xeon(R) Gold 6248 CPU at 2.50GHz with $188$GB of RAM.
Our \emph{code and datasets} are available at
\url{https://github.com/lorrainea/LAMAS}.

\subsection{Correlation with ED}\label{accuracy}

\paragraph{Task and data.} We evaluate how closely \textsc{LAMAS}, \textsc{MS}, and $q$-gram distance
correlate with ED, and how their correlations with ED 
vary with \emph{divergence} (i.e., the number of edit operations applied to a sequence, divided by its length). For \textsc{LAMAS} and \textsc{MS}, we convert match lengths into
symmetric ACS-based dissimilarities; see~\cite{DBLP:journals/jcb/UlitskyBTC06}.
We use $400$ pairs of DNA sequences of length $n\approx 100{,}000$, generated with the \textsf{generate\_dataset} tool of the Wavefront alignment algorithm~\cite{WFA}, which simulates genomic variability: $100$ pairs at each of the divergence levels $15\%$, $20\%$, $25\%$, and $30\%$.

\begin{figure}[t]
    \centering
    
    \begin{subfigure}{0.24\textwidth}
        \centering
        \includegraphics[width=\linewidth]{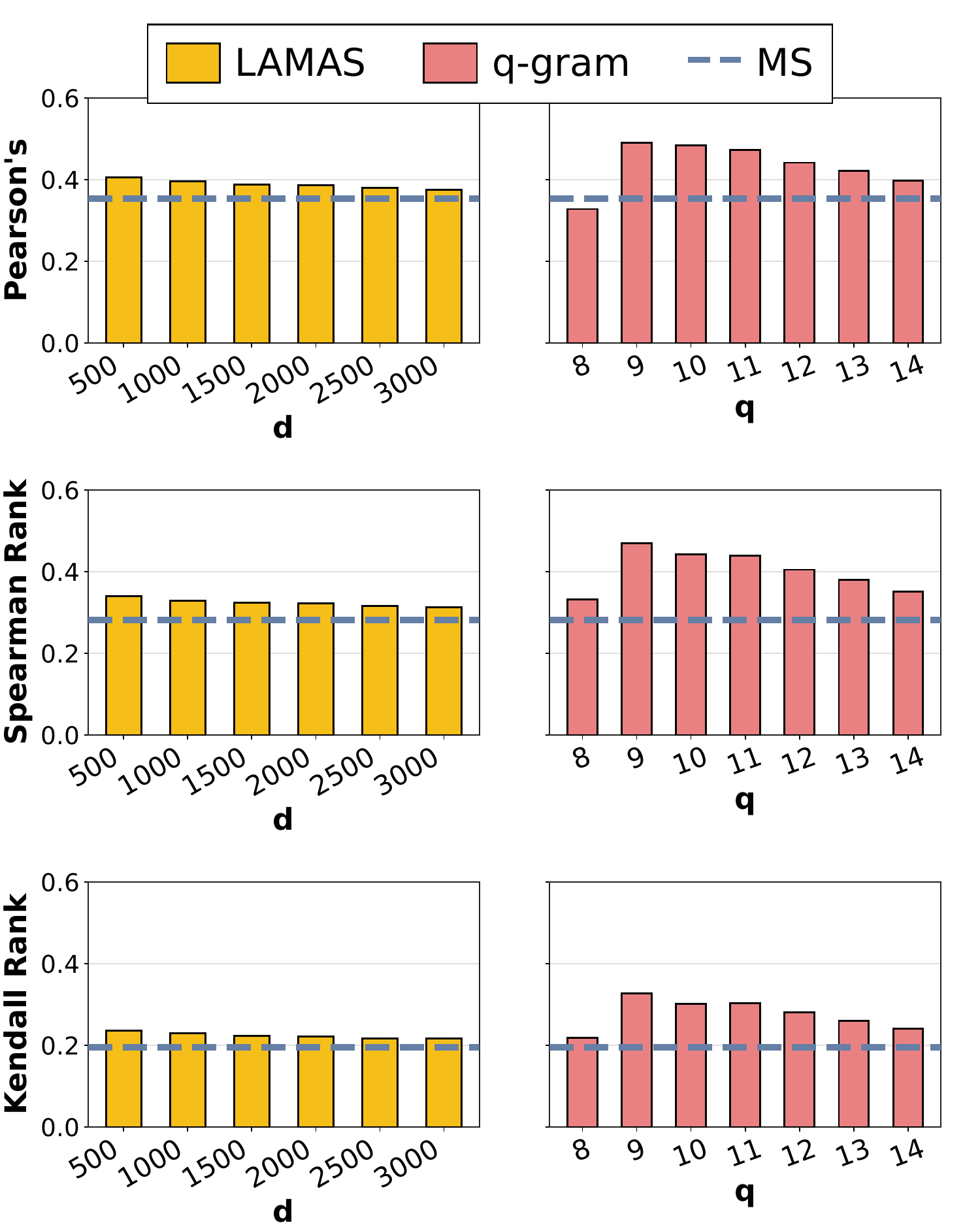}
        \caption{}
        \label{fig:100000.15}
    \end{subfigure}
    \begin{subfigure}{0.24\textwidth}
        \centering
        \includegraphics[width=\linewidth]{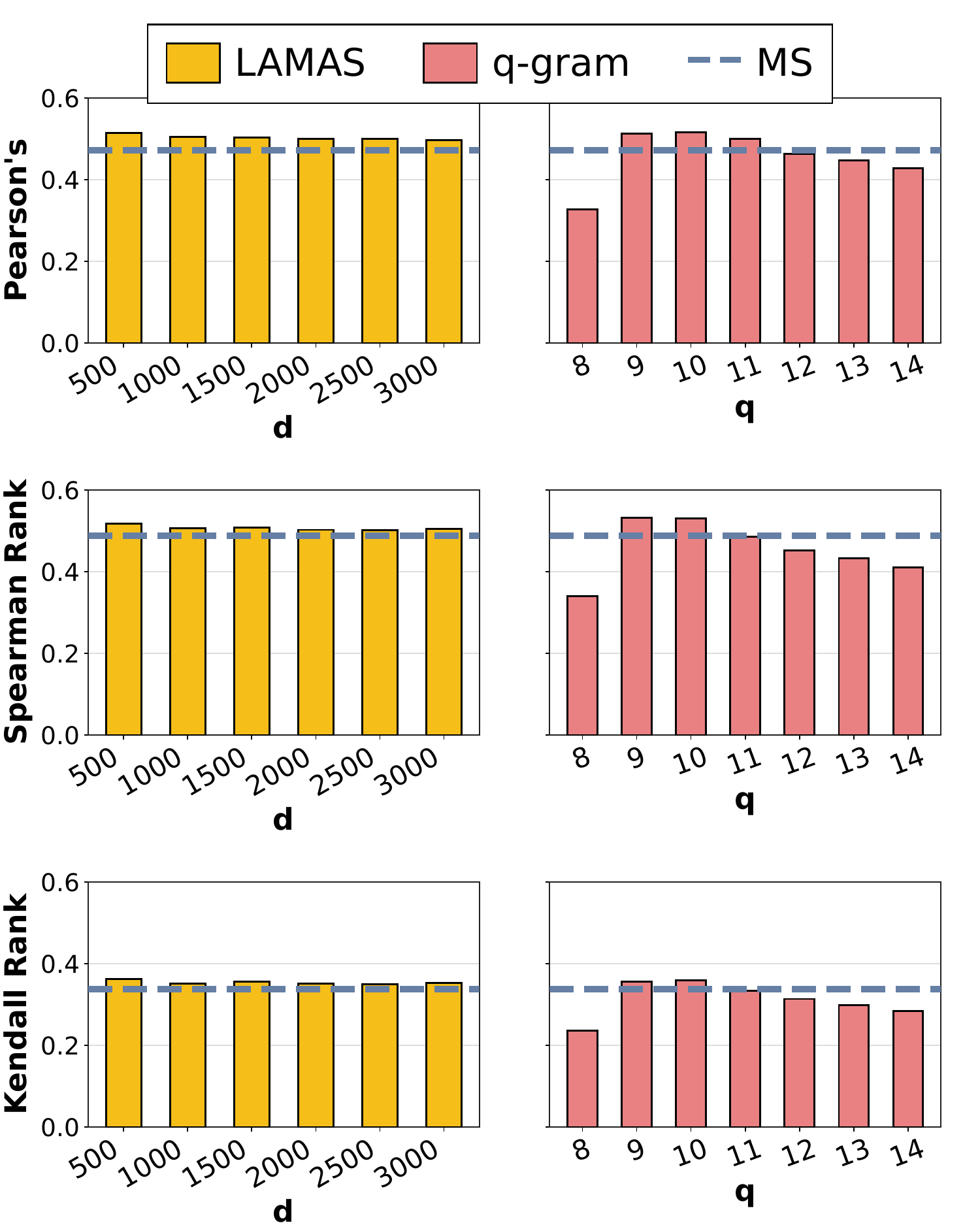}
        \caption{}
        \label{fig:100000.20}
    \end{subfigure}
    \begin{subfigure}{0.24\textwidth}
        \centering
        \includegraphics[width=\linewidth]{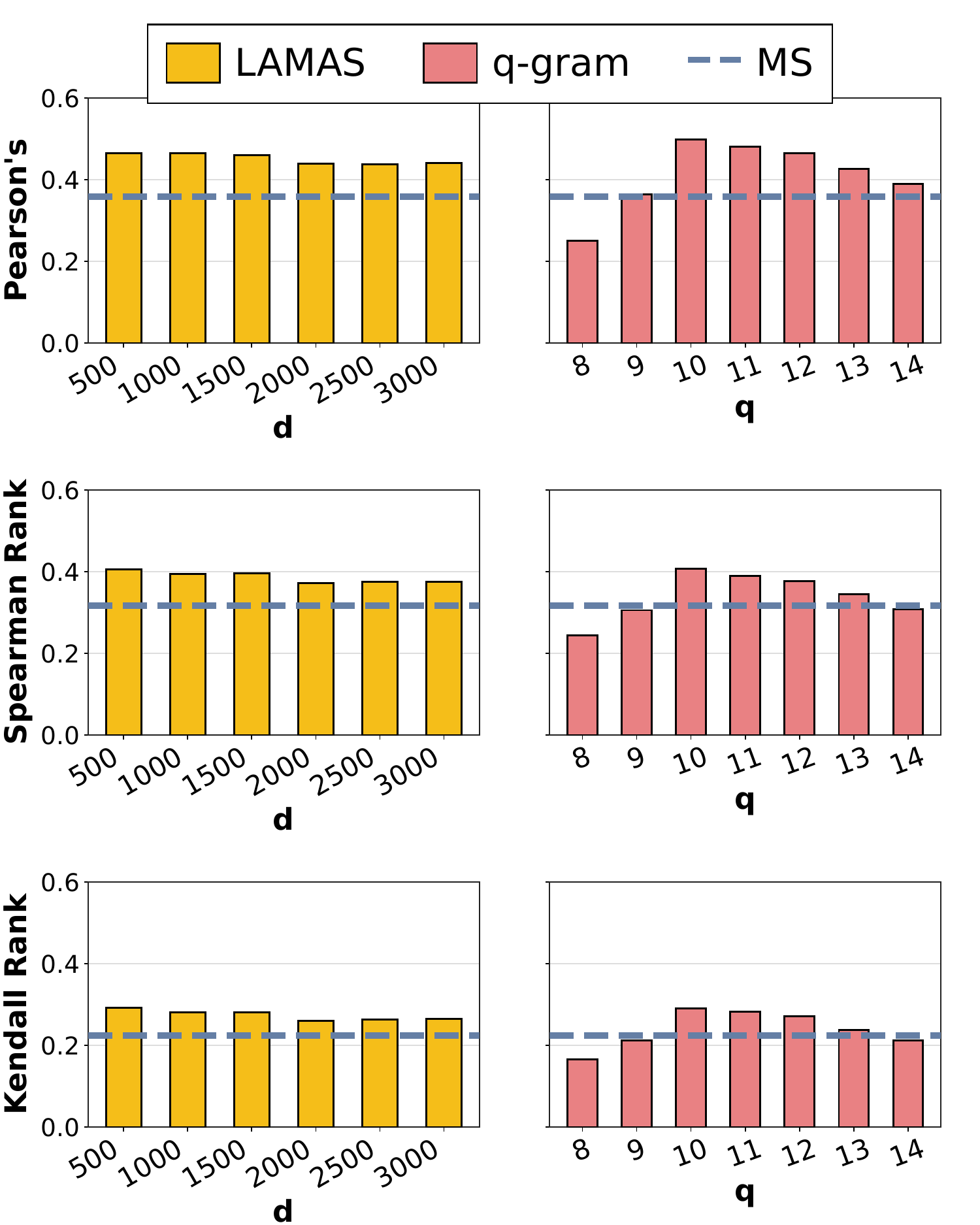}
        \caption{}
        \label{fig:100000.25}
    \end{subfigure}
    \begin{subfigure}{0.24\textwidth}
        \centering
        \includegraphics[width=\linewidth]{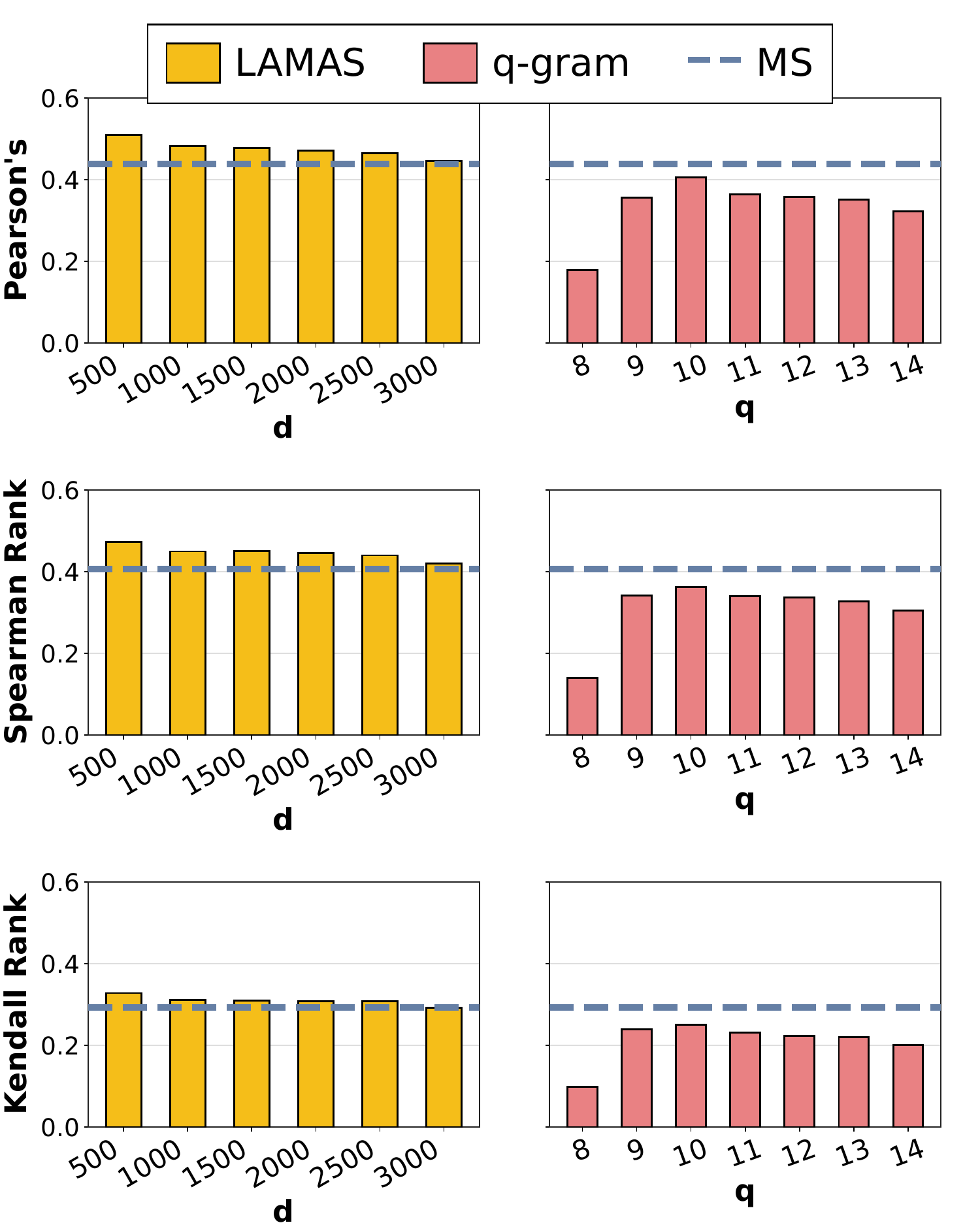}
        \caption{}
        \label{fig:100000.30}
    \end{subfigure}

    \caption{Pearson's, Spearman Rank and Kendall Rank correlation between \textsc{LAMAS} and ED, \textsc{MS} and ED and $q$-gram distance and ED, for $100$ pairs of sequences of length $n \approx 100,000$ and a divergence of: (a) $15\%$, (b) $20\%$, (c) $25\%$, and (d) $30\%$.}
    \label{fig:correlation}
\end{figure}

\paragraph{Measures.}
We compute \textsc{LAMAS} with varying $d\in\{500,1000,\ldots,3000\}$,
$q$-gram distance with varying $q\in\{8,9, \ldots,14\}$, and \textsc{MS}.  
We assess their correlation with ED using three well-known correlation measures:  Pearson's~\cite{pearson},
Spearman's~\cite{spearman}, and Kendall's Rank~\cite{kendall} correlation. 

\paragraph{Results.}\Cref{fig:correlation} shows that \textsc{LAMAS} 
\emph{achieves substantially stronger correlations with} \textsc{ED} \emph{than} \textsc{MS}; larger by $12.4\%$ on average and up to $30.8\%$.  
\textsc{LAMAS} also compares favorably to $q$-gram distance: 
averaged over all tested parameter values and divergence
levels,  it achieves $11.3\%$, $10.6\%$, and $12.2\%$ higher Pearson's, Spearman's, and Kendall's Rank  correlation, 
respectively. 
Its advantage increases at $30\%$
divergence, where these improvements are up to  $42.4\%$, 
$45.1\%$, and $48.3\%$, respectively. 
\textsc{LAMAS} outperforms even the best $q$-gram distance configuration in this divergence level by $26\%$--$31\%$. 

\subsection{Pair Classification}\label{pair_classification}\label{sec:pair-classification}

\paragraph{Task and data.} We evaluate whether the locations of shared substrings help
distinguish time series from the same class from those belonging
to different classes. We use the 15 real datasets in
\Cref{tab:lamas-pair-classification}, each containing $N$
labeled time series. We discretize each time series of every dataset into a string of length $n$, as explained below.
We construct pairs of strings and call a pair \emph{positive}
if the source time series have the same class label, and
\emph{negative} otherwise. For example, in the PenDigits~\cite{ds_pendigits} dataset, two recordings of the same digit form a positive pair. 
Then, we compute
a similarity score for the pair using one of the tested measures, as explained below, and compare it with a
validation-selected threshold to predict whether the pair is
positive. Class labels establish the ground truth, for
parameter selection and evaluation.

\paragraph{From time series to strings.} 
We convert each numeric dimension into a  string using
standard SAX preprocessing~\cite{DBLP:journals/datamine/LinKWL07}: $z$-normalization, 
overlap-weighted piecewise aggregate
approximation (PAA)~\cite{keogh}, and discretization using
Gaussian breakpoints. 
If all time series in a dataset have length $L$, we set the string length to $n=\min\{128,L\}$. Otherwise, we use $n=128$; overlap-weighted aggregation maps series with  length less than $128$ to $128$ symbols. 
For each dataset, we select each method's alphabet size $\sigma\in\{2,4,6,8,12,16\}$, parameters, and decision threshold by maximizing validation $F_1$~\cite{DBLP:books/daglib/0021593}. This allows each method to use its best validation-selected configuration. We report $F_1$ on the test pairs using the selected settings. 

\begin{table}[t]
\centering
\caption{Datasets and positive-class test $F_1$.
$n$ is SAX length and $N$ the total series count. 
Best (second-best) results
are bold (underlined).}
\label{tab:lamas-pair-classification}
\footnotesize
\setlength{\tabcolsep}{2pt}
\begin{tabular*}{\textwidth}
{@{\extracolsep{\fill}}lrr rrr rr rrr rr r@{}}
\toprule
 & & & \multicolumn{3}{c}{\textsc{LAMAS}}
 & \multicolumn{2}{c}{\textsc{MS}}
 & \multicolumn{3}{c}{$q$-gram}
 & \multicolumn{2}{c}{ED}
 & banded DTW \\
\cmidrule(lr){4-6}\cmidrule(lr){7-8}
\cmidrule(lr){9-11}\cmidrule(lr){12-13}\cmidrule(lr){14-14}
Dataset & $n$ & $N$
 & $F_1$ & $\sigma$ & $d$
 & $F_1$ & $\sigma$
 & $F_1$ & $\sigma$ & $q$
 & $F_1$ & $\sigma$
 & $F_1$ \\
\midrule
EOGHorizontalSignal~\cite{ds_eog} & 128 & 724 & \textbf{0.7383} & 6 & 19 & 0.6958 & 8 & 0.6956 & 6 & 4 & \underline{0.7362} & 6 & 0.7102 \\
Chinatown~\cite{ds_chinatown} & 24 & 363 & \textbf{0.7612} & 16 & 0 & 0.6276 & 8 & 0.7074 & 6 & 3 & \underline{0.7111} & 8 & 0.6851 \\
WordSynonyms~\cite{ds_wordsynonyms} & 128 & 905 & \textbf{0.6894} & 6 & 37 & 0.6748 & 4 & 0.6772 & 6 & 8 & \underline{0.6821} & 4 & 0.6797 \\
Meat~\cite{ds_meat} & 128 & 120 & \underline{0.7843} & 12 & 1 & 0.7442 & 8 & \textbf{0.7880} & 12 & 8 & 0.7840 & 12 & 0.7789 \\
ACSF1~\cite{ds_acsf1} & 128 & 200 & \underline{0.7348} & 4 & 20 & 0.7244 & 4 & \textbf{0.7521} & 2 & 6 & 0.7016 & 4 & 0.6870 \\
CharacterTrajectories~\cite{ds_character} & 128 & 2858 & \underline{0.8840} & 6 & 10 & 0.7809 & 16 & 0.7905 & 4 & 8 & 0.8726 & 2 & \textbf{0.8991} \\
ArticularyWordRecognition~\cite{ds_articulary} & 128 & 575 & 0.8852 & 8 & 12 & 0.8134 & 16 & 0.8415 & 16 & 2 & \underline{0.9218} & 8 & \textbf{0.9220} \\
GesturePebbleZ1~\cite{ds_pebble} & 128 & 304 & \underline{0.8609} & 4 & 9 & 0.7248 & 12 & 0.7540 & 4 & 1 & \textbf{0.8951} & 4 & 0.7699 \\
UWaveGestureLibrary~\cite{ds_uwave} & 128 & 440 & 0.7718 & 4 & 16 & 0.6927 & 2 & 0.6327 & 4 & 4 & \textbf{0.8169} & 2 & \underline{0.8050} \\
PenDigits~\cite{ds_pendigits} & 8 & 10992 & \underline{0.7524} & 2 & 1 & 0.7145 & 2 & 0.7053 & 2 & 4 & 0.7440 & 4 & \textbf{0.7999} \\
FacesUCR~\cite{ds_faces} & 128 & 2250 & 0.6892 & 6 & 5 & 0.6378 & 6 & 0.6688 & 4 & 3 & \textbf{0.7383} & 6 & \underline{0.7294} \\
ShakeGestureWiimoteZ~\cite{ds_shake} & 128 & 100 & 0.8101 & 6 & 21 & 0.6748 & 8 & 0.5484 & 12 & 1 & \underline{0.8210} & 8 & \textbf{0.8407} \\
Fish~\cite{ds_fish} & 128 & 350 & 0.6846 & 16 & 6 & 0.6670 & 16 & \underline{0.6900} & 8 & 1 & \textbf{0.7507} & 8 & 0.6786 \\
AconityMINIPrinterSmall~\cite{ds_aconity} & 128 & 881 & \underline{0.7012} & 8 & 0 & 0.6660 & 8 & 0.6651 & 4 & 1 & 0.6840 & 8 & \textbf{0.7195} \\
MixedShapesSmallTrain~\cite{ds_mixedshapes} & 128 & 2525 & \underline{0.7184} & 12 & 13 & 0.6569 & 16 & 0.6299 & 16 & 8 & \textbf{0.7203} & 12 & 0.7096 \\
\midrule
Mean $\pm$ SD & -- & --
 & \shortstack{\underline{0.7644}\\$\pm 0.0688$} & -- & --
 & \shortstack{0.6997\\$\pm 0.0516$} & --
 & \shortstack{0.7031\\$\pm 0.0741$} & -- & --
 & \shortstack{\textbf{0.7720}\\$\pm 0.0770$} & --
 & \shortstack{0.7610\\$\pm 0.0792$} \\
\bottomrule
\end{tabular*}
\end{table}

\paragraph{From matches to a pair decision.}
We summarize the matching lengths using five statistics.
For each $k\in\{1,2,3,5,8\}$, we define $f_k(X,Y)$ as
the average of the fraction of positions in $Y$ with
matching length at least $k$ in $X$ and the corresponding
fraction with $X$ and $Y$ exchanged.
For each statistic, we standardize $f_k(X,Y)$ using
its mean $\mu_k$ and standard deviation $s_k$ estimated
from the fitting pairs, and apply
$\tanh((f_k(X,Y)-\mu_k)/(2s_k))$ to bound its contribution
to $(-1,1)$. Tanh-based normalization has been used
for combining matching scores~\cite{JainNR05}. 
The same scoring procedure is used for LAMAS and MS. 

We classify each pair using a
validation-selected threshold. For multivariate
time series, scores are averaged across dimensions.
ED and $q$-gram distances are converted into
similarity scores. 
As a baseline, we apply
banded DTW~\cite{RatanamahatanaKeogh2004} to normalized PAA
sequences before SAX, selecting its warping band
and decision threshold on validation pairs.

\paragraph{Parameter selection and evaluation.} 
We retain the training/test split of each dataset and divide
the training strings into fitting (60\%) and validation
(40\%) subsets, preserving class proportions.
Within each subset, we construct balanced positive/negative
pairs, using up to $2000$ fitting, $2000$  validation, and
$20000$ test pairs. Fitting pairs determine feature
standardization; validation pairs select each method's
alphabet size and threshold, as well as \textsc{LAMAS}'s
$d\in\{0,\ldots,n-1\}$ and the baseline's
$q\in\{1,2,3,4,6,8\}$. Banded DTW's radius and threshold are selected by maximizing validation $F_1$. 
All configurations use the same subsets and pairs, with
no sequence shared across subsets.
We evaluate the selected configurations on test pairs
using $F_1$~\cite{DBLP:books/daglib/0021593}. Predicting
every pair positive scores $2/3$ on our balanced pairs.

\paragraph{Results.} Among the three ED proxies in \Cref{tab:lamas-pair-classification}, \textsc{LAMAS} achieves the highest mean $F_1$ ($0.7644$), outperforming \textsc{MS} and $q$-gram distance by $9.2\%$ and $8.7\%$, respectively, and approaching the effectiveness of ED ($0.7720$). Banded DTW has a lower mean $F_1$ ($0.7610$). \textsc{LAMAS} outperforms \textsc{MS} on all $15$ datasets, 
ranks first among all methods on $3$, and ranks second on $7$. A two-sided paired Wilcoxon signed-rank test across
the $15$ datasets confirms that \emph{\textsc{LAMAS}
significantly outperforms} \textsc{MS} ($p<0.001$)
and $q$-gram distance ($p<0.01$).
Its differences from ED and banded DTW are \emph{not} statistically significant ($p=0.64$ and $p=0.98$, respectively). 
The selected $d$'s are much smaller than $n$ (in MS $d=n-1$), which shows the benefit of restricting match positions in classification.

\subsection{Top-\texorpdfstring{$K$}{K} Retrieval}\label{top-k}

\paragraph{Task and data.}
We evaluate the effectiveness of \textsc{LAMAS}, \textsc{MS}, $q$-gram distance, and MinSearch~\cite{DBLP:conf/kdd/Zhang020} in retrieving the same top-$K$ most similar strings to a query string (pattern) as ED. We include MinSearch, a state-of-the-art approach for top-$K$ retrieval under ED, with probabilistic correctness guarantees.\footnote{We adapt the authors' code~\cite{MinSearchCode} to accept short strings
and safely verify ED on unpadded inputs. No exhaustive
fallback is added, so fewer than $K$ results may be returned. 
We set the internal gram length to
$q_{\mathrm{MSrch}}=\lceil\log_8 n\rceil$.
We select $\alpha\in\{0.5,1,2,3\}$, which controls the partition
levels searched, separately for each $K$ by maximizing mean
validation $F_1$@$K$.} We do not compare against the newer heuristic of~\cite{DBLP:journals/tkde/LoukidesPS23} whose advantage is efficiency, since our focus is on  effectiveness. 

We use all $15$ datasets in
\Cref{tab:retrieval-fixed}. From the datasets that contain multivariate timeseries, we retain one dimension uniformly at random per timeseries. 
For each dataset, we search its original training
sequences using $10$ validation and $30$ disjoint
test queries sampled from its test partition. All methods use the same sequences and queries. Sequences are $z$-normalized
and converted to SAX strings using PAA 
and Gaussian breakpoints~\cite{DBLP:journals/datamine/LinKWL07}. All methods use $\sigma=8$.   

\begin{table}[t]
\centering
\caption{Direct retrieval $F_1$@10. LAMAS and MS use the same five-statistic score. $n$ is the SAX length and
$N$ the total series count. Best (second-best) scores are bold
(underlined).}
\label{tab:retrieval-fixed}
\footnotesize
\setlength{\tabcolsep}{2.5pt}
\begin{tabular*}{\textwidth}{@{\extracolsep{\fill}}lrrrcrrcr@{}}
\toprule
Dataset & $n$ & $N$ & \textsc{LAMAS} & $d$ &
\textsc{MS} & $q$-gram & $q$ &
\shortstack{MinSearch} \\
\midrule
EOGHorizontalSignal~\cite{ds_eog} & 128 & 724 & \underline{0.8500} & $5$ & 0.3133 & 0.4467 & $8$ & \textbf{0.9167} \\
Chinatown~\cite{ds_chinatown} & 24 & 363 & \textbf{0.9033} & $1$ & 0.8200 & \underline{0.8233} & $3$ & 0.3740 \\
WordSynonyms~\cite{ds_wordsynonyms} & 128 & 905 & \underline{0.6467} & $10$ & 0.2167 & 0.2867 & $4$ & \textbf{0.6567} \\
Meat~\cite{ds_meat} & 128 & 120 & \textbf{1.0000} & $0$ & 0.9467 & 0.9533 & $4$ & \underline{0.9900} \\
ACSF1~\cite{ds_acsf1} & 128 & 200 & \textbf{0.9367} & $1$ & 0.5567 & 0.5500 & $4$ & \underline{0.9200} \\
CharacterTrajectories~\cite{ds_character} & 128 & 2858 & \underline{0.5467} & $5$ & 0.1333 & 0.2667 & $4$ & \textbf{0.8867} \\
ArticularyWordRecognition~\cite{ds_articulary} & 128 & 575 & \textbf{0.6567} & $5$ & 0.2033 & 0.1967 & $8$ & \underline{0.5333} \\
GesturePebbleZ1~\cite{ds_pebble} & 128 & 304 & \textbf{0.7600} & $8$ & 0.4467 & 0.4600 & $3$ & \underline{0.5104} \\
UWaveGestureLibrary~\cite{ds_uwave} & 128 & 440 & \textbf{0.7400} & $15$ & 0.3033 & 0.3967 & $4$ & \underline{0.6861} \\
PenDigits~\cite{ds_pendigits} & 8 & 10992 & \textbf{0.9800} & $0$ & \underline{0.8267} & \underline{0.8267} & $3$ & 0.6606 \\
FacesUCR~\cite{ds_faces} & 128 & 2250 & \textbf{0.6033} & $6$ & 0.2567 & \underline{0.3000} & $4$ & 0.2595 \\
ShakeGestureWiimoteZ~\cite{ds_shake} & 128 & 100 & \textbf{0.7300} & $5$ & 0.4500 & \underline{0.4900} & $4$ & 0.3213 \\
Fish~\cite{ds_fish} & 128 & 350 & \underline{0.5867} & $3$ & 0.3567 & 0.4867 & $8$ & \textbf{0.9767} \\
AconityMINIPrinterSmall~\cite{ds_aconity} & 128 & 881 & \textbf{0.6000} & $7$ & 0.1233 & 0.2300 & $3$ & \underline{0.4370} \\
MixedShapesSmallTrain~\cite{ds_mixedshapes} & 128 & 2525 & \textbf{0.7333} & $10$ & 0.4933 & 0.4667 & $4$ & \underline{0.6733} \\
\midrule
Mean $\pm$ SD
& --- & ---
& \shortstack{\textbf{0.7516}\\$\pm 0.1502$}
& ---
& \shortstack{0.4298\\$\pm 0.2593$}
& \shortstack{0.4787\\$\pm 0.2289$}
& ---
& \shortstack{\underline{0.6535}\\$\pm 0.2448$} \\
\bottomrule
\end{tabular*}
\end{table}

\paragraph{Scoring and parameter selection.}
For \textsc{MS} and \textsc{LAMAS}, we use the same
five-statistic similarity score described in
\Cref{pair_classification}, with normalization based on
unlabeled database pairs. The $q$-gram baseline ranks by
the $\ell_1$ distance between substring-count vectors.
We retain \textsc{LAMAS}'s displacement tolerance 
$d\in\{0,\ldots,n-1\}$ and the $q$-gram length
$q\in\{1,2,3,4,6,8\}$ previously selected by maximizing
validation $F_1$@10; these choices remain fixed across $K$. 

\paragraph{Evaluation and results.}
ED supplies the ground truth. 
Let $r$ be the $K$th smallest distance,
$L$ the records closer than $r$, $E$ those at distance $r$,
and $R$ the returned records. $F_1@K=2c/(|R|+K)$, where 
$c=|R\cap L|+\min\{|R\cap E|,K-|L|\}$. Thus, boundary-tied records are interchangeable and missing closer records is penalized. We average $F_1$@$K$ over test queries within each dataset and then across datasets with equal weight.

\begin{figure}[!ht]
    \centering
    \includegraphics[width=0.6\linewidth]{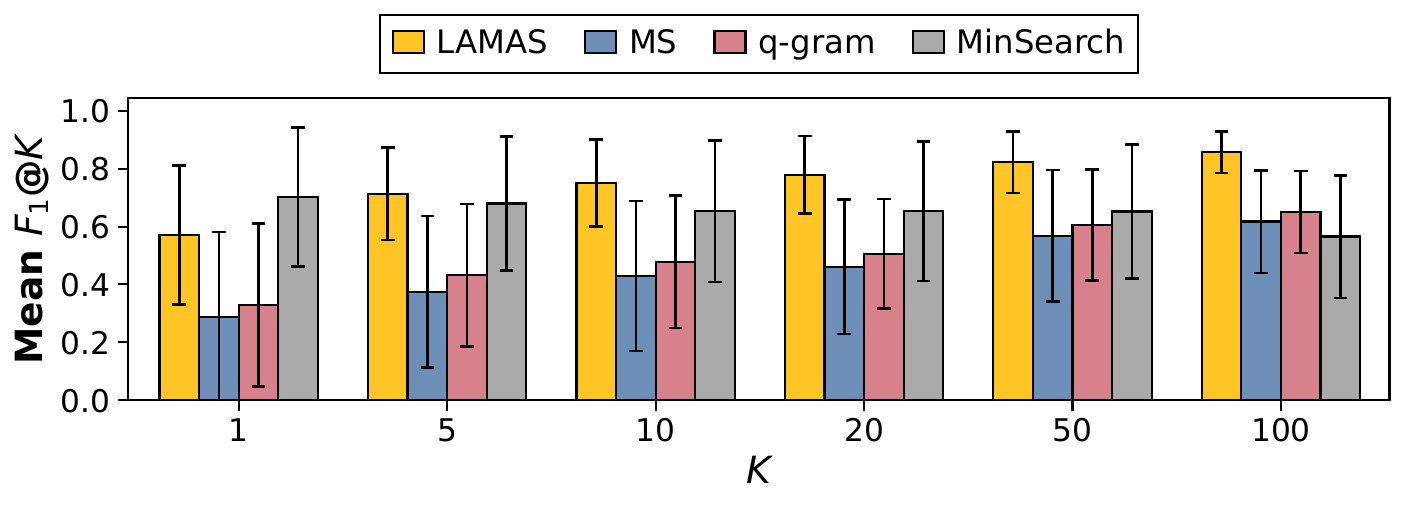}
    \caption{Mean $F_1$@$K$ vs. $K$; 
error bars show the sample standard deviation; datasets with size $\leq K$ are excluded.}
    \label{fig:topK_vs_K}
\end{figure}

\Cref{tab:retrieval-fixed} shows that \textsc{LAMAS} 
outperforms both \textsc{MS} and $q$-gram distance on all $15$
datasets at $K=10$ (Wilcoxon signed-rank tests across datasets,
$p<0.001$ for both). The improvement over MS is $74.9\%$. \textsc{LAMAS} 
also outperforms MinSearch on $11$ out of $15$ datasets,
although the difference is not statistically significant
($p=0.107$). \Cref{fig:topK_vs_K} reports sensitivity to $K$. 
\textsc{LAMAS} outperforms both \textsc{MS} and $q$-gram
distance at every evaluated $K$, with lower variation across
datasets.  It also outperforms MinSearch on average for
$K\geq5$ (e.g., its $F_1$@100 is $0.8565$ vs. $0.5656$).  
The values of $d$ are smaller than $n$, showing the effectiveness of restricting match positions in \textsc{LAMAS} compared to \textsc{MS} ($d=n-1$).  

\subsection{Clustering}\label{clustering}

\paragraph{Tasks.} We evaluate clustering quality using ARI~\cite{HubertArabie1985}, 
NMI~\cite{StrehlGhosh2002}, 
purity~\cite{DBLP:books/daglib/0021593},  
and class-support-weighted
$F_1$~\cite{sklearnF1}, 
computed after optimally matching clusters to
reference classes using a $1$-to-$1$  assignment; higher values are better, ARI lies in $[-0.5,1]$ and all other measures in $[0,1]$. 

We use real datasets and smaller real-data subsets selected for clear class separation under ED. 
\begin{table}[!ht]
  \centering
  \caption{Clustering datasets and selected parameters:
  $N$ strings of length $n$, alphabet size $\sigma$,
  and $k$ classes. Parameters $d$ and $q$ are selected
  on training data for the first two datasets and
  on the evaluated subsets for the remaining three.}
  \label{tab:real_datasets}
  \footnotesize
  \setlength{\tabcolsep}{3pt}
  \begin{tabular}{lrrrrrr}
    \toprule
    Dataset & $N$ & $n$ & $\sigma$ & $k$ & $d$ & $q$ \\
    \midrule
    \textsf{Crop}~\cite{lamasCropDataset}
      & 16,800 & 128 & 8 & 24 & 8 & 6 \\
    \textsf{StarLightCurves}~\cite{lamasStarLightDataset}
      & 8,236 & 128 & 8 & 3 & 16 & 8 \\
    \midrule
    \textsf{CinCECGTorso}~\cite{Dau2019UCR}
      & 235 & 512 & 8 & 4 & 24 & 4 \\
    \textsf{ECG5000}~\cite{Dau2019UCR}
      & 408 & 140 & 8 & 5 & 26 & 2 \\
    \textsf{ADFA-LD}~\cite{CreechHu2013ADFA}
      & 24 & 500 & $\approx170$ & 2 & 3 & 6 \\
    \bottomrule
  \end{tabular}
\end{table}

\paragraph{Real datasets.}
We show that \textsc{LAMAS} can \emph{cluster real datasets substantially more effectively} than \textsc{MS}
and $q$-gram distance, while remaining competitive with ED. We 
use two real datasets (see \Cref{tab:real_datasets}): \textsf{Crop}, containing image-derived time series of
length $L=46$, and \textsf{StarLightCurves}, containing
stellar-brightness series of length $L=1024$.
We cluster their test partitions and select parameters
using training data. Specifically, we use complete-linkage agglomerative clustering~\cite{clusteringbook},
cutting each dendrogram into $k$ clusters, where $k$ is the number 
of reference classes (\Cref{tab:real_datasets}).
For \textsc{LAMAS} and \textsc{MS}, dissimilarity is $1-\overline{\ell}/n$,
where $\overline{\ell}$ is the mean match length over both
directions; $q$-gram distance is the $\ell_1$ difference of
substring-count vectors, including overlaps.

Both datasets undergo Gaussian smoothing with standard
deviation $\max(1,0.01L)$, $z$-normalization,
overlap-weighted PAA~\cite{keogh}, and SAX discretization using
standard-normal breakpoints~\cite{DBLP:journals/datamine/LinKWL07}.
We use $n=128$ and $\sigma=8$; for \textsf{Crop}, this increases the
representation length without adding information.
A fixed uniform sample of $500$ training sequences per dataset,
drawn without inspecting labels or distances, is used to select
parameters by maximizing ARI over 
$d\in\{0,1,2,4,8,16,32,64\}$ and $q\in\{1,2,3,4,6,8\}$,
breaking ties toward smaller values. 

\begin{figure}[t]
    \centering
    \begin{subfigure}[t]{0.42\linewidth}
        \centering
        \includegraphics[width=\linewidth]{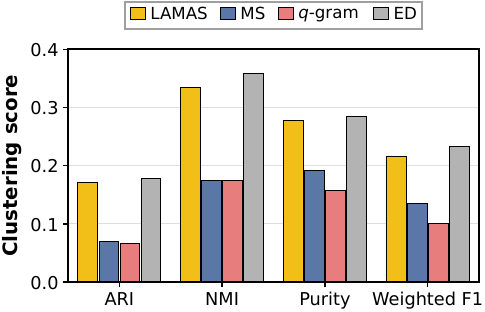}
        \caption{}
        \label{fig:crop-clustering}
    \end{subfigure}
    \hfill
    \begin{subfigure}[t]{0.42\linewidth}
        \centering
        \includegraphics[width=\linewidth]{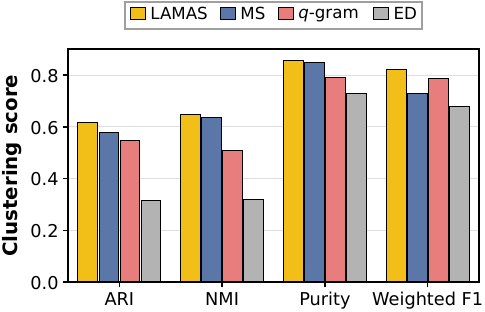}
        \caption{}
        \label{fig:starlight-clustering}
    \end{subfigure}
    \caption{Clustering performance on the complete test partitions
    of (a) \textsf{Crop} and
    (b) \textsf{StarLightCurves}.} 
    \label{fig:clustering-full}
\end{figure}

\textsc{LAMAS} outperforms \textsc{MS} and $q$-grams on all four measures (\Cref{fig:clustering-full}). For \textsf{Crop}, its ARI is $146\%$ higher than \textsc{MS} and $154\%$ higher than $q$-grams, while ED performs
slightly better. 
For \textsf{StarLightCurves}, \textsc{LAMAS} ranks first on all four
measures, improving ARI and weighted $F_1$ by
$6.8\%$ and $12.8\%$ over \textsc{MS}, and by
$13.1\%$ and $4.5\%$ over $q$-grams, respectively; its ARI is also $96\%$ higher than that of ED. 
The selected bounds $d=8,16$ are far smaller than $n=128$,
whereas \textsc{MS} corresponds to $d=127$.
\begin{figure}[t]
\centering
\begin{subfigure}[t]{0.32\linewidth}
    \centering
    \includegraphics[width=\linewidth]{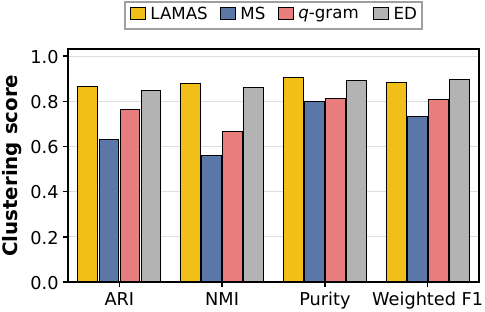}
    \caption{}
\end{subfigure}\hfill
\begin{subfigure}[t]{0.32\linewidth}
    \centering
    \includegraphics[width=\linewidth]{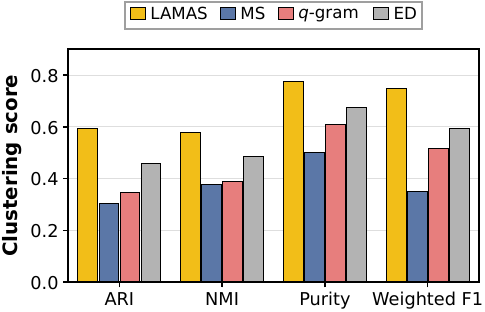}
    \caption{}
\end{subfigure}\hfill
\begin{subfigure}[t]{0.32\linewidth}
    \centering
    \includegraphics[width=\linewidth]{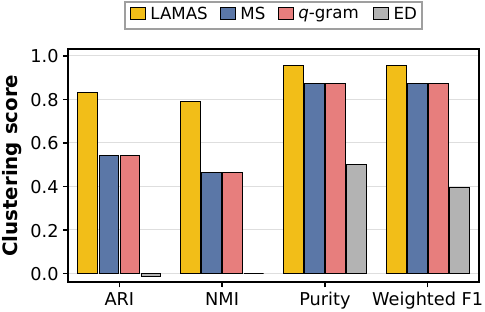}
    \caption{}
    \label{fig:clustering-adfa}
\end{subfigure}
\caption{Clustering of (a) \textsf{CinCECGTorso}, (b) \textsf{ECG5000}, and (c) \textsf{ADFA-LD}.}
\label{fig:clustering-three}
\end{figure}

\paragraph{ED-selected case studies.}
We show that \textsc{LAMAS} \emph{recovers reference-class structure much better} than
\textsc{MS} and $q$-gram distance, often outperforming ED itself, even
on collections selected to favor class separation under ED.  
We use three real datasets (see \Cref{tab:real_datasets}):
\textsf{CinCECGTorso}, \textsf{ECG5000}, and %
the system-call
dataset \textsf{ADFA-LD}. 
For \textsf{CinCECGTorso} and \textsf{ECG5000}, we use the same normalization, PAA,
and SAX discretization as above, without Gaussian smoothing.  
For \textsf{ADFA-LD}, we collapse consecutive repeated calls,
concatenate traces within each class using unique boundary
tokens, and extract length-$500$ windows with stride $250$;
system calls are the alphabet letters. 
After constructing these strings, we compute each class medoid
under ED and retain strings with the largest difference between
their distance to the nearest other-class medoid and their
distance to their own-class medoid. 
The resulting subset sizes are given in \Cref{tab:real_datasets}.
Reference labels guide subset selection, tuning of $d$ and $q$ on the selected subsets  
and evaluation; they are not inputs to the pairwise
string-comparison scores or the clustering merge decisions. 

As can be seen in \Cref{fig:clustering-three}, \textsc{LAMAS} outperforms both \textsc{MS} and $q$-gram distance 
and also outperforms 
ED in nearly all comparisons,
despite the subsets being selected to favor class separation
under ED. For example, \textsc{LAMAS} improves ARI and weighted
$F_1$ over ED by approximately $31\%$ and $26\%$ on
\textsf{ECG5000}, and improves ARI and NMI over the
strongest competitor by $53\%$ and $70\%$ on \textsf{ADFA-LD}.

\subsection{Time and Space Efficiency}\label{efficiency}

\paragraph{Real DNA sequencing data.}
We use two datasets of Oxford Nanopore Technologies (ONT)
\emph{reads} (sequenced DNA fragments): \textsf{ONT1}, with read lengths
$40{,}000$--$50{,}000$ and $12\%$ divergence~\cite{50k_dataset};
and \textsf{ONT2}, with $50$ pairs of ultra-long reads of length
$500{,}135$--$848{,}888$, mapped to the CHM13 v1.1 human reference
genome, with average divergence $6.1\%$~\cite{500k_dataset}.
For both datasets, we set $q=4$ and
$d=\lfloor0.1\cdot \min(|X|,|Y|)\rfloor$ for each string  pair $(X,Y)$. We compare LAMAS to its non-optimized versions and to the string-based distance measures. 

\begin{figure}[t]
    \centering
    \includegraphics[width=\linewidth]{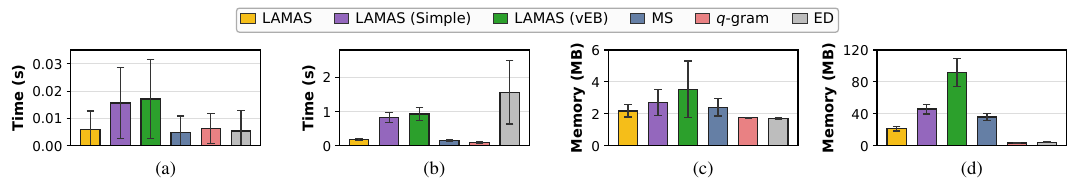}
    \caption{Running time in seconds for (a) \textsf{ONT1} and (b) \textsf{ONT2}. Peak memory in MB for (c) \textsf{ONT1} and (d) \textsf{ONT2}.}
    \label{fig:500k}
\end{figure}

\Cref{fig:500k} shows that \textsc{LAMAS} takes $11\%$ longer than ED on \textsf{ONT1} reads, but is $8.5$ times faster on the much longer 
\textsf{ONT2} reads. \textsc{LAMAS} is $61\%$ and $65\%$ faster than
\textsc{LAMAS} (\textsc{Simple}) and \textsc{LAMAS}
(\textsc{vEB}), respectively, on \textsf{ONT1},
and $78\%$ and $80\%$ faster on \textsf{ONT2}. 
Its running time is comparable to $q$-gram distance  on \textsf{ONT1} but larger in \textsf{ONT2}.  
\textsc{LAMAS} is $27\%$ and $26\%$ slower than \textsc{MS}
on \textsf{ONT1} and \textsf{ONT2}, respectively, a modest
overhead for enforcing the displacement constraint that
improves effectiveness in our experiments.

\Cref{fig:500k} shows that \textsc{LAMAS} uses $9\%$
and $41\%$ less memory than \textsc{MS} on \textsf{ONT1}
and \textsf{ONT2}, respectively.
It also uses $20\%$ and $38\%$ less memory than
\textsc{LAMAS} (\textsc{Simple}) and \textsc{LAMAS}
(\textsc{vEB}), respectively, on \textsf{ONT1},
and $54\%$ and $77\%$ less on \textsf{ONT2}. Although ED and $q$-gram distance use less memory,
\textsc{LAMAS} requires only $21.4$ MB on average
on \textsf{ONT2}, where it is $8.5$ times faster than ED. 

\paragraph{Scalability.}
We use substrings of Chromosomes 1 and 2 of the human
reference genome 
\cite{NC_000001.11,NC_000002.12}. We compare \textsc{LAMAS} with its non-optimized versions, \textsc{MS}, and $q$-gram distance. We omit ED, as it was much slower, and include banded DTW for reference, after encoding DNA letters as integers. 

\begin{figure}[t]
    \centering
    \includegraphics[width=\linewidth]{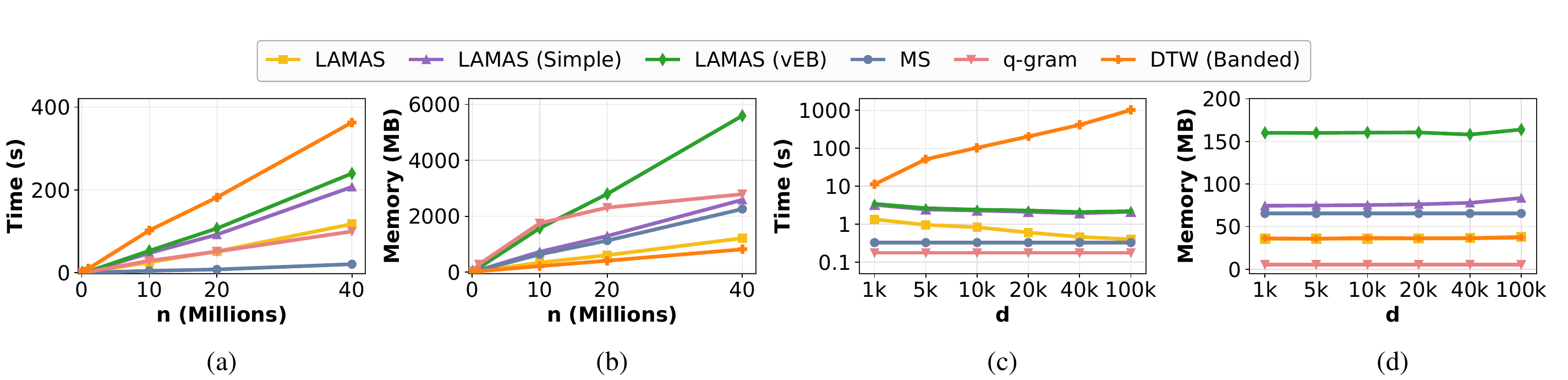}
    \caption{(a) Running time in seconds and (b) peak memory in MB, for varying $n$. (c) Running time in seconds and (d) peak memory in MB, for varying $d$.}
    \label{fig:varying_n_d}
\end{figure}

\emph{Impact of $n$.}~\Cref{fig:varying_n_d}(a) shows that the running times
grow linearly with $n$.
We fix $d=1000$, the DTW band radius $d=1000$, and $q=12$; a $q$ value in $[8, 16]$ is usually used for DNA sequences~\cite{Kent2002BLAT}. 
\textsc{LAMAS} consistently outperforms \textsc{LAMAS}
(\textsc{Simple}) and \textsc{LAMAS} (\textsc{vEB})
for $n\geq10^5$. At $n=4\cdot10^7$, its running
time is $43\%$ lower than that of \textsc{LAMAS}
(\textsc{Simple}) and $51\%$ lower than that of
\textsc{LAMAS} (\textsc{vEB}). \textsc{LAMAS} is on average 
$11.3\%$ faster than $q$-gram distance and $81.4\%$ faster
than banded DTW. MS is much faster than LAMAS, but it does not enforce the displacement constraint, which is useful in the above applications. 
\Cref{fig:varying_n_d}(b) shows that \textsc{LAMAS} achieves its running-time improvements while also reducing memory use.
At $n=4\cdot10^7$, it uses $53\%$ and $78\%$ less
peak memory than \textsc{LAMAS} (\textsc{Simple}) and
\textsc{LAMAS} (\textsc{vEB}), respectively.
It also uses $46\%$ less memory than \textsc{MS}
and $56\%$ less than $q$-gram distance, but $49\%$
more than banded DTW.
Thus, Improvement~$2$ in \Cref{sec:linear} improves
\emph{both} time and space over the other
\textsc{LAMAS} implementations.

\emph{Impact of $d$.}~\Cref{fig:varying_n_d}(c) ($n=10^6$) shows that
increasing $d$ from $1000$ to $10^5$ reduces
\textsc{LAMAS}'s running time by $70\%$.
On average over the six values of $d$, \textsc{LAMAS} is
$68.6\%$, $70.5\%$, and $97.5\%$ faster than \textsc{LAMAS}
(\textsc{Simple}), \textsc{LAMAS} (\textsc{vEB}), and banded DTW,
respectively. $q$-gram distance is fastest, followed by
\textsc{MS}, which at $d=10^5$ is only $22\%$ faster than \textsc{LAMAS}.
\Cref{fig:varying_n_d}(d) shows that \textsc{LAMAS}
uses $52\%$--$54\%$, $77\%$--$78\%$, and $42\%$--$45\%$ less memory
than \textsc{LAMAS} (\textsc{Simple}), \textsc{LAMAS}
(\textsc{vEB}), and \textsc{MS}, respectively; it is comparable to banded DTW,
and $q$-gram distance uses the least.

\section{Conclusion}\label{sec:conclusion}

We introduced LAMAS, a \emph{location-aware} generalization of the well-known MS primitive~\cite{rw:ChangLawler94}, and developed a time-optimal algorithm for computing it that requires $\cO(n)$ time and space. Our experiments show that LAMAS improves similarity-based classification, retrieval, and clustering. The implementation of 
our LAMAS algorithm scales linearly in $n$ 
and does not slow down as the displacement tolerance grows.

\section*{Declarations}

The conceptualization, key ideas, and research design originated from the authors. We used GPT-6 Astra and Claude Fable 5.1 to improve the paper's presentation (including the design of figures and plots), optimize the source code, and assist with the preparation of experiments. The authors remain fully responsible for the correctness and originality of the content, the references, and for the interpretation of the results. 

\bibliographystyle{alphaurl}
\bibliography{references}

\end{document}

%% file: algorithm.tex
\section{LAMAS in Linear Time}\label{sec:algo}

We first present a simple $\cO(n\log d)$-time algorithm for LAMAS and then improve it to $\cO(n)$ time. 
Our improvements are related to the way queries are answered. 

Throughout, we set $S:=X\#Y$ (the string of \Cref{sec:prelim}), so that $|S|=|X|+|Y|+1\le 2n+1$.
Without loss of generality, we assume $1\le d<n$: for $d=0$, we have
$\ell_i=\LCP(X[i\dd|X|),Y[i\dd|Y|))$ and $\delta_i=0$ for every $i<|X|$ (and $\ell_i=0$
otherwise), which is one LCE query per position; for $d\ge n$, the output is the same as
for $d=n-1$, since every $j\in[0\dd|X|)$ satisfies $|j-i|\le n-1$.

\subsection{The Simple Algorithm}\label{sec:simple}

\paragraph{Main idea.}
For each position $i$ in $Y$, the algorithm considers
the suffixes of $X$ starting within distance $d$ of $i$.
These starting positions form a sliding window
$[i-d\dd i+d]$, restricted to the valid positions of $X$.
As $i$ increases, the window moves one position to the right.
By \Cref{fact:lcp-min}, the suffix sharing the longest
prefix with $Y[i\dd|Y|)$ is either its lexicographic
predecessor or successor among the admissible suffixes.
The algorithm computes the LCP length of both candidates and reports the larger one as $\ell_i$ (breaking ties arbitrarily), together with the displacement $\delta_i=j^*-i$ of the selected suffix $X[j^*\dd|X|)$.
Thus, after constructing the suffix array of $S=X\#Y$,
LAMAS reduces to predecessor and successor queries
among the suffix-array ranks of a sliding window.
These ranks are maintained in a balanced search tree (e.g., an AVL tree), 
supporting queries and updates in $\cO(\log d)$ time. As the window slides $\cO(n)$ times in total, the algorithm takes $\cO(n \log d)$ time. 

\Cref{fig:simple-idea} illustrates this idea for $X=\texttt{banana}$, $Y=\texttt{anana}$, and $d=1$ at $i=1$: the window $\{0,1,2\}$ has ranks $\{6,3,9\}$, and the query suffix $Y[1\dd|Y|)=\texttt{nana}$ has rank $8$, whose predecessor $6$ ($\texttt{banana}$) and successor $9$ ($\texttt{nana}$) have LCP lengths $0$ and $4$; hence, $j^*=2$ and $(\ell_1,\delta_1)=(4,1)$.

\begin{figure}[t]
\centering
\resizebox{0.55\linewidth}{!}{%
\begin{tikzpicture}[
    x=1cm,y=1cm,
    font=\sffamily\scriptsize,
    box/.style={rounded corners=2pt,minimum width=0.72cm,
                minimum height=0.54cm,inner sep=0pt,line width=0.7pt},
    active/.style={draw=green!55!black,fill=green!11},
    added/.style={draw=green!65!black,fill=green!30,line width=1.0pt},
    inactive/.style={draw=black!22,fill=black!5},
    query/.style={draw=blue!65!black,fill=blue!11},
    existing/.style={fill=green!60!black},
    qdot/.style={fill=blue!65!black}
]
    \node[anchor=west,font=\sffamily\bfseries\small] at (0.1,4.33)
        {(a) Sliding window and query suffix};

    \node[anchor=east,font=\sffamily\bfseries] at (0.79,3.15) {$X$:};
    \foreach \p/\letter/\rank in {
        0/b/6,1/a/3,2/n/9,
        3/a/4,4/n/10,5/a/5
    } {
        \pgfmathsetmacro{\xx}{1.30+1.05*\p}
        \node[font=\sffamily\scriptsize,text=black!70]
            at (\xx,3.64) {\p};
        \ifnum\p<2
            \node[box,active,font=\ttfamily\large] at (\xx,3.15) {\letter};
            \node[text=green!55!black] at (\xx,2.70) {\rank};
        \else
            \ifnum\p=2
                \node[box,added,font=\ttfamily\large] at (\xx,3.15) {\letter};
                \node[text=green!55!black] at (\xx,2.70) {\rank};
            \else
                \node[box,inactive,font=\ttfamily\large] at (\xx,3.15) {\letter};
                \node[text=black!60] at (\xx,2.70) {\rank};
            \fi
        \fi
    }
    \node[anchor=east,text=black!60] at (0.80,2.70) {rank:};
    \node[anchor=west,text=green!45!black] at (0.83,2.17)
        {$i=1$: admissible ranks $\{6,3,9\}$};
    \draw[->,line width=0.7pt,green!55!black]
        (3.20,3.86) -- (3.20,3.49);
    \node[anchor=south,text=green!45!black]
        at (3.40,3.80) {add $j=2$};

    \node[anchor=east,font=\sffamily\bfseries] at (0.79,1.36) {$Y$:};
    \foreach \p/\letter in {
        0/a,1/n,2/a,3/n,4/a
    } {
        \pgfmathsetmacro{\xx}{1.30+1.05*\p}
        \ifnum\p=0
            \node[box,inactive,font=\ttfamily\large] at (\xx,1.36) {\letter};
        \else
            \node[box,query,font=\ttfamily\large] at (\xx,1.36) {\letter};
        \fi
    }
    \node[anchor=west,text=blue!55!black] at (0.83,0.80)
        {$Y[1\dd|Y|)=\texttt{nana}$ has rank $8$};

    \draw[black!22,line width=0.5pt] (0.12,0.46)--(7.72,0.46);

    \node[anchor=west,font=\sffamily\bfseries\small] at (0.1,0.13)
        {(b) Predecessor and successor among window ranks};

    \draw[black!45,line width=0.6pt] (0.70,-0.89)--(7.47,-0.89);
    \foreach \r in {0,...,11} {
        \pgfmathsetmacro{\xx}{0.72+0.60*\r}
        \draw[black!40,line width=0.6pt]
            (\xx,-0.80)--(\xx,-0.98);
        \node[text=black!60] at (\xx,-1.22) {\r};
    }
    \foreach \r in {3,6,9} {
        \pgfmathsetmacro{\xx}{0.72+0.60*\r}
        \fill[green!60!black] (\xx,-0.89) circle (0.115);
    }
    \fill[blue!65!black] (5.52,-0.89) circle (0.138);
    \node[anchor=south,text=blue!60!black] at (5.52,-0.57)
        {query rank $8$};

    \node[anchor=west,text=green!45!black] at (0.76,-1.68)
        {predecessor $6$: LCP $0$};
    \node[anchor=east,text=green!45!black] at (7.44,-1.68)
        {successor $9$: LCP $4$};
    \node[anchor=west,font=\sffamily\bfseries] at (0.76,-2.18)
        {Selected $j^*=2$; output $(\ell_1,\delta_1)=(4,1)$};
\end{tikzpicture}%
}
\caption{Main idea of the simple algorithm:
(a) moving the window to $i=1$ adds position $2$ of $X$, of rank $9$;
(b) the predecessor and successor of the query rank $8$ have LCP
lengths $0$ and $4$, respectively.}
\label{fig:simple-idea}
\end{figure}

\paragraph{The algorithm.}
\begin{algorithm}[t]\small
\caption{Simple algorithm for LAMAS}
\label{alg:lamas-simple}
\begin{algorithmic}[1]
\Require Strings $X,Y$ and displacement tolerance $d$
\Ensure Arrays $\ell$ and $\delta$

\State $S \gets X\#Y$ \label{ln:simple-step1-start}
\State Compute $\textsf{SA}_S$, $\textsf{iSA}_S$,
       and the LCE structure of $S$
       \label{ln:simple-step1-end}

\State $A[j] \gets \textsf{iSA}_S[j]$
       for $j\in[0\dd|X|)$
       \label{ln:simple-step2-start}
\State $B[i] \gets \textsf{iSA}_S[|X|+1+i]$
       for $i\in[0\dd|Y|)$
       \label{ln:simple-step2-end}

\State Initialize an empty AVL tree $W$
       \label{ln:simple-step3-start}
\For{$j=0$ \textbf{to} $\min(d,|X|-1)$}
    \State $W.\Call{Insert}{A[j]}$
\EndFor

\For{$i=0$ \textbf{to} $|Y|-1$}
    \If{$i>0$}
        \If{$i+d<|X|$}
            \State $W.\Call{Insert}{A[i+d]}$
        \EndIf
        \If{$i>d$}
            \State $W.\Call{Delete}{A[i-d-1]}$
        \EndIf
    \EndIf

    \State $r_p \gets W.\Call{Predecessor}{B[i]}$
    \State $r_s \gets W.\Call{Successor}{B[i]}$
           \label{ln:simple-step3-end}

    \State $(\ell_i,\delta_i)\gets(0,\bot)$
           \label{ln:simple-step4-start}
    \For{$r\in\{r_p,r_s\}\setminus\{\bot\}$}
        \State $j\gets\textsf{SA}_S[r]$
        \State $L\gets\Call{LCE}{j,|X|+1+i}$
        \If{$L>\ell_i$ \textbf{or} $\delta_i=\bot$}
            \State $(\ell_i,\delta_i)\gets(L,j-i)$
        \EndIf
    \EndFor
    \label{ln:simple-step4-end}
\EndFor

\State \Return $(\ell,\delta)$
\end{algorithmic}
\end{algorithm}
\Cref{alg:lamas-simple} presents the simple algorithm,
which consists of four steps: 
(1) It computes $\textsf{SA}_S$, $\textsf{iSA}_S$,
and the LCE structure of $S$ (\Cref{sec:prelim}),
allowing constant-time LCP queries between suffixes of
$X$ and $Y$
(lines~\ref{ln:simple-step1-start}--\ref{ln:simple-step1-end}).
(2) It constructs arrays $A$ and $B$ containing
the ranks of the suffixes of $X$ and $Y$, respectively,
in $\textsf{SA}_S$
(lines~\ref{ln:simple-step2-start}--\ref{ln:simple-step2-end}).  Since the suffix of $S$ starting with $\#$ has the
    largest rank, the values of $A$ and $B$ are pairwise distinct and form exactly the set
    $[0\dd|A|+|B|)$. 
(3) For each position $i$, it computes the predecessor
$r_p$ and successor $r_s$ of $B[i]$ among the ranks
in the window
\[
W_i:=\{A[j]:j\in [0\dd|X|),\ |j-i|\le d\}.
\]
To this end, it maintains $W_i$, which has at most $2d+1$ elements, in an AVL tree updated as the window slides (lines~\ref{ln:simple-step3-start}--\ref{ln:simple-step3-end}).
(4) It uses at most two LCE queries, one for each
candidate, and reports the maximum LCP length $\ell_i$
and the corresponding displacement $\delta_i=j^*-i$,
where $j^*$ is the starting position of a maximizing
suffix of $X$
(lines~\ref{ln:simple-step4-start}--\ref{ln:simple-step4-end}).

Correctness follows from \Cref{fact:lcp-min}: the ranks in $W_i$ are exactly the
ranks of the admissible suffixes $X[j\dd|X|)$ with $|j-i|\le d$, and, as noted in
\Cref{sec:prelim}, the LCP with $Y[i\dd|Y|)$ is maximized by the nearest rank below or
above $B[i]$. Steps~1, 2, and~4 take $\cO(n)$ time and space. Step~3 performs $\cO(n)$
AVL tree operations on a tree of at most $2d+1$ elements, i.e., $\cO(\log d)$ time
each, and uses $\cO(d)$ space. Thus, we obtain: 

\begin{proposition}\label{prop:simple}
The LAMAS problem can be solved in $\cO(n \log d)$ time using $\cO(n)$ space.
\end{proposition}

\begin{example}[Simple Algorithm]\label{ex:simple}
Let $X=\texttt{banana}$, $Y=\texttt{anana}$, and $d=1$.
Steps~1--2 compute the suffix arrays of $S=X\#Y$
(\Cref{fig:example-sa}) and the rank arrays
$A=(6,3,9,4,10,5)$ and $B=(2,8,1,7,0)$
(\Cref{fig:example-ab}). 
For $i=0$, $W_0=\{6,3\}$ contains no predecessor of $B[0]=2$.
Its successor, rank $3$, corresponds to $X[1\dd|X|)=\texttt{anana}$,
yielding $(\ell_0,\delta_0)=(5,1)$.
For $i=1$, inserting $A[2]=9$ gives $W_1=\{6,3,9\}$
(\Cref{fig:simple-idea}). The predecessor $6$ and successor $9$
of $B[1]=8$ give LCP lengths $0$ and $4$, respectively,
yielding $(\ell_1,\delta_1)=(4,1)$. 
For $i=2$, removing $A[0]=6$ and inserting $A[3]=4$
gives $W_2=\{3,9,4\}$. The successor of $B[2]=1$
is rank $3$, corresponding to $j=1$, yielding
$(\ell_2,\delta_2)=(3,-1)$. Although $j=3$ also attains
length $3$, its rank $4$ need not be examined: any maximizing
position suffices. \Cref{fig:example-lamas} shows the complete execution.
\end{example}

\begin{figure}[t]
\centering
\footnotesize
\begin{subfigure}{\linewidth}
\centering
\setlength{\tabcolsep}{3.4pt}
\begin{tabular}{l|cccccccccccc}
\toprule
$t$ & $0$ & $1$ & $2$ & $3$ & $4$ & $5$ & $6$ & $7$ & $8$ & $9$ & $10$ & $11$ \\
\midrule
$S[t]$ & \texttt{b} & \texttt{a} & \texttt{n} & \texttt{a} & \texttt{n} & \texttt{a} & $\#$ & \texttt{a} & \texttt{n} & \texttt{a} & \texttt{n} & \texttt{a} \\
$\textsf{SA}_S[t]$  & $11$ & $9$ & $7$ & $1$ & $3$ & $5$ & $0$ & $10$ & $8$ & $2$ & $4$ & $6$ \\
$\textsf{iSA}_S[t]$ & $6$ & $3$ & $9$ & $4$ & $10$ & $5$ & $11$ & $2$ & $8$ & $1$ & $7$ & $0$ \\
\bottomrule
\end{tabular}
\caption{Step~1: $S=X\#Y$ with its suffix array and inverse suffix array.}
\label{fig:example-sa}
\end{subfigure}

\medskip
\begin{subfigure}{\linewidth}
\centering
\setlength{\tabcolsep}{5pt}
\begin{tabular}{l|cccccc}
\toprule
$j$ (resp.~$i$) & $0$ & $1$ & $2$ & $3$ & $4$ & $5$ \\
\midrule
$A[j]=\textsf{iSA}_S[j]$   & $6$ & $3$ & $9$ & $4$ & $10$ & $5$ \\
$B[i]=\textsf{iSA}_S[7+i]$ & $2$ & $8$ & $1$ & $7$ & $0$ & -- \\
\bottomrule
\end{tabular}
\caption{Step~2: ranks $A$ of the suffixes of $X$ and $B$ of the suffixes of $Y$.}
\label{fig:example-ab}
\end{subfigure}

\medskip
\begin{subfigure}{\linewidth}
\centering
\setlength{\tabcolsep}{4pt}
\begin{tabular}{l|ccccc}
\toprule
$i$ & $0$ & $1$ & $2$ & $3$ & $4$ \\
\midrule
$B[i]$   & $2$ & $8$ & $1$ & $7$ & $0$ \\
$W_i$    & $\{6,3\}$ & $\{6,3,9\}$ & $\{3,9,4\}$ & $\{9,4,10\}$ & $\{4,10,5\}$ \\
$r_p$, $r_s$ & $\bot$, $3$ & $6$, $9$ & $\bot$, $3$ & $4$, $9$ & $\bot$, $4$ \\
$\ell_p$, $\ell_s$ & $-\infty$, $5$ & $0$, $4$ & $-\infty$, $3$ & $0$, $2$ & $-\infty$, $1$ \\
$\ell_i$ & $5$ & $4$ & $3$ & $2$ & $1$ \\
$j^*$    & $1$ & $2$ & $1$ & $2$ & $3$ \\
$\delta_i$ & $1$ & $1$ & $-1$ & $-1$ & $-1$ \\
\bottomrule
\end{tabular}
\caption{Steps~3 and~4: windows, nearest ranks $r_p$, $r_s$, their LCP lengths $\ell_p$, $\ell_s$ with $Y[i\dd|Y|)$, and output ($\bot$ and $-\infty$ denote undefined).}
\label{fig:example-lamas}
\end{subfigure}
\caption{The simple algorithm on $X=\texttt{banana}$, $Y=\texttt{anana}$, and $d=1$ (\Cref{ex:simple}).}
\label{fig:example}
\end{figure}

\subsection{Improvement \texorpdfstring{$1$}{1}: Universe Reduction}\label{sec:improved}

\paragraph{Main idea.}
The $\cO(\log d)$ cost per operation of the algorithm in \Cref{sec:simple} comes from comparison-based
predecessor search. A vEB tree answers predecessor queries in $\cO(\log\log u)$ time
over a universe of size $u$. However, our ranks range over $[0\dd|S|)$, making $u$ too large.
Our \emph{key observation} is that the windows of $d$ consecutive query positions involve at most 
$3d$ consecutive entries of $A$ and $d$ query keys from $B$. Since the queries only depend on the relative order of keys, we can replace the global ranks of keys with \emph{local} ranks without changing the answers. We therefore cut the problem into $\cO(n/d)$ independent
blocks, each with at most $4d$ keys, and \emph{re-rank} the keys of every block to $[0\dd 4d)$ by a 
single global radix sort; this %
makes the
universe small enough for vEB trees. Each sliding window can then be processed in
$\cO(\log\log d)$ time using a vEB tree,
yielding $\cO(d\log\log d)$ time per block and
$\cO(n\log\log d)$ time overall. 

\paragraph{An auxiliary problem.}
We isolate the task of finding the predecessor and successor
of each query suffix rank within its admissible window by defining and solving \Cref{pro:swps}.

\begin{problem}[Sliding-Window Predecessor/Successor (SWPS)]\label{pro:swps}
Given a positive integer $N$ and arrays
$a=(a_0,\ldots,a_{3N-1})$ and $b=(b_0,\ldots,b_{N-1})$
of pairwise distinct integers in $[0\dd 4N)$,\footnote{Some entries
of $a$ may be undefined, and $b$ may have fewer than $N$ entries, to handle the boundaries of $X$ and $Y$;
undefined entries are ignored.} 
output, for every $i\in[0\dd N)$, the predecessor and successor
of $b_i$ in the window $\{a_j : j\in[i\dd i+2N]\}$; an answer is
reported as undefined if no such element exists.
\end{problem}

\begin{proposition}\label{lem:correct-core}
    The SWPS problem can be solved in $\cO(N\log\log N)$ time using
    $\cO(N)$ space.
\end{proposition}
\begin{proof}
Initialize a vEB tree over $[0\dd 4N)$; insert
the defined values among $a_0,\ldots,a_{2N}$ into it. This takes $\cO(N)$ time and space.
For $i\in[0\dd N)$, we query the tree for the predecessor and the successor of
$b_i$, and then remove $a_i$ and insert $a_{i+2N+1}$ (if $i+2N+1<3N$; undefined entries are
skipped), so that the tree contains exactly the window of $b_{i+1}$. Each operation takes
$\cO(\log\log N)$ time, and there are $\cO(N)$ of them.
\end{proof}

\paragraph{The algorithm.}
Steps~1, 2, and~4 of the simple algorithm remain unchanged; we replace Step~3 by a
reduction to $\cO(n/d)$ instances of SWPS. 

The first step of the reduction is to partition $B$ into blocks
$B_k:=B[kd\dd(k+1)d)$ of $d$ positions, for $k\in[0\dd\lceil|Y|/d\rceil)$, and pair every
$B_k$ with the block $A_k:=A[(k-1)d\dd(k+2)d)$ of $3d$ entries of $A$. This is exactly
the SWPS structure: the window of $B[kd+i']$, for $i'\in [0\dd d)$, is
$A[(k-1)d+i'\dd(k-1)d+i'+2d]$, i.e., the positions $[i'\dd i'+2d]$ of $A_k$. Entries of
$A_k$ outside $[0\dd|X|)$ are undefined, and the last block of $B$ may be shorter than $d$.
Consecutive blocks of $A$ overlap by $2d$ positions, so every entry of $A$ belongs to at
most three blocks and the total size of all instances is $\cO(n)$.

The second step is to reduce the universe of each instance to $[0\dd 4d)$ to be able to apply \Cref{lem:correct-core}. 
We form, for every $k$, the set $M_k$ of the (defined)
values of $A_k$ and $B_k$, each tagged with its block $k$ and its index in $A_k$ or $B_k$. We sort all $M_k$'s \emph{together} by one radix sort: a counting sort of all tags by 
value, over the universe $[0\dd|S|)$, followed by a stable counting sort by block index.
This takes $\cO(n)$ time. Replacing every value by its position in the sorted $M_k$ yields
arrays $A'_k$ and $B'_k$ over $[0\dd 4d)$ that have the same relative order as $A_k$ and
$B_k$; hence predecessors and successors are preserved. 

Finally, we solve each SWPS instance
$a:=A'_k$, $b:=B'_k$, $N:=d$ by \Cref{lem:correct-core}, in $\cO(d\log\log d)$ time, reusing
(and re-initializing, in $\cO(d)$ time) a single vEB tree for all instances. 

\emph{Complexity.}~The total time is $\cO(n)$ for Steps~1, 2, and~4 and for the radix sort, plus
$\cO(n/d)\cdot\cO(d\log\log d)=\cO(n\log\log d)$ for the SWPS instances. If
$d=\log^{\cO(1)}n$, we replace the vEB tree by a dynamic fusion
tree~\cite{DBLP:conf/focs/PatrascuT14}, which supports every operation on at most $2d+1$
keys in $\cO(\log_w d)=\cO(\log d/\log w)\subseteq\cO(\log d/\log\log n)=\cO(1)$ time, where $w=\Omega(\log n)$ is
the word size in bits.

\begin{proposition}\label{prop:improved}
The LAMAS problem can be solved in $\cO(n \log \log d)$ time using $\cO(n)$ space.
If $d=\log^{\cO(1)} n$, then it  can be solved in $\cO(n)$ time using $\cO(n)$ space.
\end{proposition}

\subsection{Improvement \texorpdfstring{$2$}{2}: Window Decomposition and Union-Find}\label{sec:linear}

\paragraph{Main idea.}We remove the dependency on $d$, showing that each SWPS instance can be solved in $\cO(N)$ time, and thus LAMAS in the optimal $\cO(n)$ time. We observe that the windows of SWPS have a rigid structure. Partition the input array $a$ into three equal parts,
(left) $a_L$, (middle) $a_M$, and (right) $a_R$.  
For each query key $b_i$ in $b$, its window consists of 
a \emph{suffix} of $a_L$, 
the \emph{entire} $a_M$, and a \emph{prefix}
of $a_R$, as illustrated in
\Cref{fig:swps-decomposition}. The middle part $a_M$ is static, so a single sorted scan answers all queries on it. The suffix of $a_L$ only shrinks as $i$ increases, and the prefix of $a_R$ only shrinks if we process the queries in \emph{decreasing} order of $i$.
Thus, all remaining queries are predecessor/successor queries on a sorted sequence that
only undergoes \emph{deletions}, which is the interval union-find problem admitting
a linear-time solution.

\begin{figure}[t]
\centering
\resizebox{0.85\linewidth}{!}{%
\begin{tikzpicture}[x=.37cm,y=.72cm,font=\sffamily\small]

\colorlet{swpsLeft}{blue!65}
\colorlet{swpsMiddle}{green!55!black}
\colorlet{swpsRight}{orange!85}
\colorlet{swpsOff}{gray!15}

\foreach \x/\partname/\partindices/\parthue in {
    0/$a_L$/$a_0\ldots a_5$/swpsLeft,
    6/$a_M$/$a_6\ldots a_{11}$/swpsMiddle,
    12/$a_R$/$a_{12}\ldots a_{17}$/swpsRight%
}{
    \node[font=\bfseries,text=\parthue]
        at (\x+3,1.85) {\partname};
    \node[font=\scriptsize,text=gray!75!black]
        at (\x+3,1.38) {\partindices};
}

\node[anchor=east] at (-.55,.45) {$W_1$ for $b_1$};
\node[anchor=east] at (-.55,-.85) {$W_4$ for $b_4$};

\foreach \index/\y in {1/0,4/-1.30}{
    \foreach \j in {0,...,17}{
        \pgfmathtruncatemacro{\active}{
            (\j>=\index && \j<=\index+12) ? 1 : 0
        }
        \ifnum\active=1
            \ifnum\j<6
                \def\thiscolor{swpsLeft}
            \else
                \ifnum\j<12
                    \def\thiscolor{swpsMiddle}
                \else
                    \def\thiscolor{swpsRight}
                \fi
            \fi
        \else
            \def\thiscolor{swpsOff}
        \fi
        \fill[\thiscolor] (\j,\y) rectangle ++(1,.78);
        \draw[white,line width=.6pt]
            (\j,\y) rectangle ++(1,.78);
    }
    \draw[gray!50,line width=.45pt]
        (0,\y) rectangle (18,\y+.78);
    \draw[gray!80,line width=.85pt]
        (6,\y)--(6,\y+.78);
    \draw[gray!80,line width=.85pt]
        (12,\y)--(12,\y+.78);
}

\node[anchor=west,font=\scriptsize,text=gray!70!black]
    at (0,-1.75)
    {Colored entries belong to $W_i$; grey entries are excluded.};

\draw[-{Stealth[length=2mm]},thick,gray!65]
    (18.9,.62)--(18.9,-.76);

\node[anchor=west,font=\scriptsize,align=left]
    at (19.2,.28) {$i$ increases};

\node[anchor=west,font=\scriptsize,align=left]
    at (19.2,-.40)
    {left suffix shrinks\\right prefix grows};

\node[anchor=west,font=\scriptsize,align=left,text=swpsRight]
    at (19.2,-1.16)
    {Reverse order:\\right prefix shrinks};

\end{tikzpicture}%
}
\caption{Decomposition of SWPS windows for $N=6$.}
\label{fig:swps-decomposition}
\end{figure}

Since $i \in[0\dd N)$, every window of $b_i$  decomposes as
\[
[i\dd i+2N]=[i\dd N)\cup[N\dd 2N)\cup[2N\dd 2N+i].
\]
Namely, let $a_L:=a_0,\ldots,a_{N-1}$, $a_M:=a_N,\ldots,a_{2N-1}$, and $a_R:=a_{2N},\ldots,a_{3N-1}$. 
Denoting by $\text{pred}_L(i)$, $\text{pred}_M(i)$, and $\text{pred}_R(i)$ the predecessor of $b_i$ in $\{a_i,\ldots,a_{N-1}\}$, $\{a_N,\ldots,a_{2N-1}\}$, and $\{a_{2N},\ldots,a_{2N+i}\}$, respectively (and analogously for $\text{succ}$), we have
\[
\text{pred}(i)=\max\{\text{pred}_L(i),\text{pred}_M(i),\text{pred}_R(i)\},
\]
\[
\text{succ}(i)=\min\{\text{succ}_L(i),\text{succ}_M(i),\text{succ}_R(i)\},
\]
where undefined values are ignored. We answer the three kinds of queries using the following tool.

\begin{lemma}[Interval Union-Find~\cite{DBLP:journals/jcss/GabowT85}]\label{lem:iuf}
Let $e_0,\ldots,e_{m-1}$ be a sequence of elements, all initially \emph{present}. There is a data structure that supports the following operations: $\textsf{delete}(k)$, which marks $e_k$ as deleted; $\textsf{next}(k)$, which returns the smallest $k'\geq k$ such that $e_{k'}$ is present (or reports that no such $k'$ exists); and $\textsf{prev}(k)$, which returns the largest $k'\leq k$ such that $e_{k'}$ is present (or reports that no such $k'$ exists). Any sequence of $q$ such operations takes $\cO(m + q)$ total time.
\end{lemma}
\begin{proof}
We maintain a partition of $[0\dd m)$ into intervals of the following form: every interval $[l\dd r]$ is such that $e_l,\ldots,e_{r-1}$ are deleted and $e_r$ is present, except possibly for the rightmost interval, whose elements may all be deleted. Initially, every element forms a singleton interval. Each interval is stored as a set in a union-find data structure, and its representative stores $l$ and $r$. Operation $\textsf{delete}(k)$ unites the interval containing $k$ with the interval containing $k+1$ (if $k<m-1$); this preserves the invariant and takes $\cO(1)$ time on top of the union operation. For $\textsf{next}(k)$, we find the interval $[l\dd r]$ containing $k$ and return $r$ if $e_r$ is present (otherwise, no such element exists). For $\textsf{prev}(k)$, we find the interval $[l\dd r]$ containing $k$; if $k=r$ and $e_r$ is present we return $k$; otherwise, $e_k$ is deleted and we return $l-1$ (if $l>0$; otherwise, no such element exists), which is present by the maximality of the intervals. Since every union merges the set containing some $k$ with the set containing $k+1$, this is the interval union-find problem: the union tree is the path with parent$(k)=k+1$, which is fixed in advance, while the deletions may arrive in any order. Any sequence of $q$ operations takes $\cO(m+q)$ total time~\cite{DBLP:journals/jcss/GabowT85}.
\end{proof}

\begin{lemma}\label{lem:swps-linear}
    The SWPS problem can be solved in $\cO(N)$ time using $\cO(N)$ space.
\end{lemma}
\begin{proof}
    As all (defined) values of $a$ and $b$ are pairwise distinct integers in $[0\dd 4N)$, we first use counting sort to obtain, in $\cO(N)$ time, the sorted sequence of all these values, where every value is annotated with its origin ($a_L$, $a_M$, $a_R$, or $b$) and its index in $a$ or $b$. A single scan over this sequence yields:
    \begin{itemize}
        \item the sorted sequences $c^L_0<c^L_1<\cdots$, $c^M_0<c^M_1<\cdots$, and $c^R_0<c^R_1<\cdots$ of the (defined) values of $a_L$, $a_M$, and $a_R$, respectively, together with the rank $\rho(j)$ of every defined $a_j$ within the sorted sequence of its own part;
        \item for every $i \in[0\dd N)$, the numbers $\pi^L_i$, $\pi^M_i$, and $\pi^R_i$ of values of $a_L$, $a_M$, and $a_R$, respectively, that are smaller than $b_i$.
    \end{itemize}
    Note that the predecessor (resp.~successor) of $b_i$ in any subset $P$ of $a_R$ is the element of $P$ with the largest rank at most $\pi^R_i-1$ (resp.~the smallest rank at least $\pi^R_i$) in $c^R$, and analogously for $a_L$ and $a_M$.

    \emph{Middle part.} Since the set $\{a_N,\ldots,a_{2N-1}\}$ is the same for all queries, we have $\text{pred}_M(i)=c^M_{\pi^M_i-1}$ (undefined if $\pi^M_i=0$) and $\text{succ}_M(i)=c^M_{\pi^M_i}$ (undefined if $\pi^M_i$ equals the number of defined values in $a_M$). This takes $\cO(1)$ time per query.

    \emph{Right part.} We process the queries in the order $i=N-1,N-2,\ldots,0$, so that the set $\{a_{2N},\ldots,a_{2N+i}\}$ only \emph{shrinks}. We initialize the data structure of \Cref{lem:iuf} over the sorted sequence $c^R$ (so $m$ is the number of defined values in $a_R$). For $i=N-1$ down to $0$, we first answer the query for $b_i$: $\text{pred}_R(i)=c^R_{\textsf{prev}(\pi^R_i-1)}$ (undefined if $\pi^R_i=0$ or $\textsf{prev}$ reports no element) and $\text{succ}_R(i)=c^R_{\textsf{next}(\pi^R_i)}$ (undefined if $\pi^R_i=m$ or $\textsf{next}$ reports no element); then, if $a_{2N+i}$ is defined, we perform $\textsf{delete}(\rho(2N+i))$. By induction, when the query for $b_i$ is answered, exactly the defined values among $a_{2N},\ldots,a_{2N+i}$ are present, as required. We perform $\cO(N)$ operations in total, which takes $\cO(N)$ time by \Cref{lem:iuf}.

    \emph{Left part.} This is symmetric: we process the queries in the order $i=0,1,\ldots,N-1$, so that the set $\{a_i,\ldots,a_{N-1}\}$ only shrinks. We initialize the data structure of \Cref{lem:iuf} over $c^L$, and, for $i=0$ to $N-1$, we answer the query for $b_i$ using $\textsf{prev}(\pi^L_i-1)$ and $\textsf{next}(\pi^L_i)$ exactly as above, and then perform $\textsf{delete}(\rho(i))$ if $a_i$ is defined. Again, this takes $\cO(N)$ time.

    We combine the partial answers into $\text{pred}(i)$ and $\text{succ}(i)$ for every $i$ in $\cO(1)$ time each to get $\cO(N)$ time and space.
\end{proof}

\paragraph{The final algorithm.}
The algorithm is that of \Cref{sec:improved}, with \Cref{lem:swps-linear} used in place of
\Cref{lem:correct-core}: Steps~1, 2, and~4 and the universe reduction take $\cO(n)$ time,
and the $\cO(n/d)$ SWPS instances take $\cO(n/d)\cdot\cO(d)=\cO(n)$ time in total. The
space is $\cO(n)$ for the LCE data structure and the sorted arrays, plus $\cO(d)=\cO(n)$
for the data structures of \Cref{lem:iuf}, which are reused across instances.

\begin{theorem}\label{the:linear}
The LAMAS problem can be solved in $\cO(n)$ time using $\cO(n)$ space, for any $d\geq 0$.
\end{theorem}

This bound is \emph{optimal}: $\Omega(n)$ time is required to read the input, and it
matches the time to compute the MS array~\cite{DBLP:conf/focs/ChangL90}.